\documentclass[10pt, conference, compsocconf]{IEEEtran}
\usepackage{cite}
\ifCLASSINFOpdf
\else
\fi
\usepackage[cmex10]{amsmath}
\usepackage{algorithmic}

\usepackage{tikz}
\usetikzlibrary{positioning}
\usetikzlibrary{quantikz}
\usepackage{subfig}
\usepackage[export]{adjustbox}
\usepackage{multirow}
\usepackage{braket}
\usepackage{listings}
\usepackage[flushleft]{threeparttable}
\usepackage[hyphens]{url}
\begin{document}
\title{Not All SWAPs Have the Same Cost: A Case for Optimization-Aware Qubit Routing}


\author{\IEEEauthorblockN{Ji Liu}
\IEEEauthorblockA{
North Carolina State University\\
Raleigh, United States\\
jliu45@ncsu.edu}\and
\IEEEauthorblockN{Peiyi Li}
\IEEEauthorblockA{
North Carolina State University\\
Raleigh, United States\\
pli11@ncsu.edu}\and
\IEEEauthorblockN{Huiyang Zhou}
\IEEEauthorblockA{
North Carolina State University\\
Raleigh, United States\\
hzhou@ncsu.edu}
}


%Telephone: (800) 555--1212, Fax: (888) 555--1212}


\maketitle
\thispagestyle{plain}

\begin{abstract}
Despite rapid advances in quantum computing technologies, the qubit connectivity limitation remains to be a critical challenge. Both near-term NISQ quantum computers and relatively long-term scalable quantum architectures do not offer full connectivity. As a result, quantum circuits may not be directly executed on quantum hardware, and a quantum compiler needs to perform qubit routing to make the circuit compatible with the device layout. During the qubit routing step, the compiler inserts SWAP gates and performs circuit transformations. Given the connectivity topology of the target hardware, there are typically multiple qubit routing candidates. The state-of-the-art compilers use a cost function to evaluate the number of SWAP gates for different routes and then select the one with the minimum number of SWAP gates. After qubit routing, the quantum compiler performs gate optimizations upon the circuit with the newly inserted SWAP gates.

In this paper, we observe that the aforementioned qubit routing is not optimal, and qubit routing should \textit{not} be independent on subsequent gate optimizations. We find that with the consideration of gate optimizations, not all of the SWAP gates have the same basis-gate cost. These insights lead to the development of our qubit routing algorithm, NASSC (Not All Swaps have the Same Cost). NASSC is the first algorithm that considers the subsequent optimizations during the routing step. Our optimization-aware qubit routing leads to better routing decisions and benefits subsequent optimizations. We also propose a new optimization-aware decomposition for the inserted SWAP gates. Our experiments show that the routing overhead compiled with our routing algorithm is reduced by up to $69.30\%$ ($21.30\%$ on average) in the number of CNOT gates and up to $43.50\%$ ($7.61\%$ on average) in the circuit depth compared with the state-of-the-art scheme, SABRE.

\end{abstract}

\begin{IEEEkeywords}
quantum computing; compiler optimization; qubit routing

\end{IEEEkeywords}

% \fi
\IEEEpeerreviewmaketitle

\section{Introduction}

Quantum computing has shown immense promise for accelerating chemistry simulation~\cite{jones2012faster}, prime factorization~\cite{shor1999shoralgorithm}, database search~\cite{grover1996grover}, and machine learning~\cite{biamonte2017qml}. Recently, Google, IBM, Intel, and Honeywell announced their quantum computers with 72, 65, 49, and 10 qubits, respectively~\cite{Google72qubit,IBM65qubit, Intel49qubit,Honeywell10qubit}. These quantum computers with few tens to hundreds of qubits are termed as Noisy Intermediate-Scale Quantum (NISQ) computers~\cite{preskill2018NISQera}. Many quantum system works~\cite{das2021jigsaw, duckering2021orchestrated, liu2021systematic, huang2021logical, li2021software, tang2021cutqc, wang2021quantumnas, liu2020quantum, gokhale2020optimized} have been recently proposed for the NISQ systems. Although NISQ computers do not have enough qubits to accommodate error correction codes, they are useful for exploring quantum algorithms and demonstrating quantum supremacy~\cite{arute2019googlesupermacy}.

Both near-term NISQ systems~\cite{arute2019googlesupermacy,IBM65qubit,Intel49qubit} and the long-term scalable quantum architectures~\cite{monroe2013scaling,monroe2014large,murali2019noise} do not support full connectivity among a high number of qubits. However, quantum algorithms are developed with an implicit assumption of a fully-connected quantum computer. Such mismatch makes the qubit mapping and routing critical challenges in quantum computing systems. 

A quantum compiler is responsible for a number of tasks, including decomposing higher-level gates to basic ones supported natively by the target quantum hardware, restructuring quantum circuits, optimizing circuits, and scheduling quantum gate operations. During the restructuring step, the compiler performs the logical-to-physical qubit mapping and qubit routing. Specifically, the compiler needs to insert SWAP gates and perform circuit transformations to make the circuit layout compatible with the device layout. When inserting the SWAP gates, the compiler would evaluate different routing candidates based on a cost function. The cost function is computed based on the number of SWAP gates~\cite{li2019sabre} and/or the fidelity of the inserted SWAP gates~\cite{niu2020hardware}. After qubit mapping and routing, the compiler performs circuit optimizations such as template matching~\cite{maslov2008gatecancellation}, commutation analysis~\cite{itoko2020mapping_commutation}, and gate cancellation~\cite{maslov2008cancellation} to optimize the circuit. In the state-of-the-art approaches~\cite{li2019sabre, niu2020hardware}, the qubit routing step selects the best route based on the backend topology, qubit fidelity, and logical circuit topology while being independent upon the subsequent circuit optimization step.

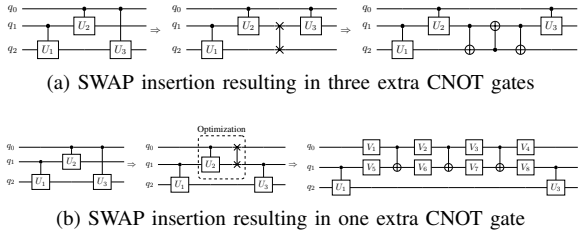
\begin{figure}[ht]
  \centering
    \subfloat[SWAP insertion resulting in three extra CNOT gates\label{subfig:threeCNOTinsertion}]{   
   \begin{adjustbox}{width=0.9\linewidth}
    \begin{quantikz}[row sep=2mm]
    \lstick{$q_0$}& \qw &\ctrl{1} & \ctrl{2} & \qw\\
    \lstick{$q_1$}&\ctrl{1} & \gate{U_2}  & \qw  & \qw\\
    \lstick{$q_2$} & \gate{U_1} & \qw& \gate{U_3} & \qw
    \end{quantikz} $\Rightarrow$ 
    \begin{quantikz}[row sep=2mm]
    \lstick{$q_0$}& \qw & \ctrl{1}&\qw & \ctrl{1} & \qw\\
    \lstick{$q_1$}&\ctrl{1} & \gate{U_2} & \swap{1}& \gate{U_3} & \qw\\
    \lstick{$q_2$} & \gate{U_1} & \qw& \targX{} &\qw & \qw
    \end{quantikz} $\Rightarrow$ 
    \begin{quantikz}[row sep=2mm]
    \lstick{$q_0$}&\qw & \ctrl{1} & \qw & \qw & \qw & \ctrl{1} & \qw\\
    \lstick{$q_1$}&\ctrl{1} & \gate{U_2} & \ctrl{1}&\targ{} & \ctrl{1} & \gate{U_3} & \qw\\
    \lstick{$q_2$} & \gate{U_1} & \qw& \targ{} &\ctrl{-1} & \targ{} & \qw & \qw
    \end{quantikz}
    
    \end{adjustbox}
    }\hfill
    
        \subfloat[SWAP insertion resulting in one extra CNOT gate\label{subfig:singleCNOTinsertion}]{   
   \begin{adjustbox}{width=0.9\linewidth}
    \begin{quantikz}[row sep=2mm]
    \lstick{$q_0$}& \qw & \ctrl{1} & \ctrl{2} & \qw\\
    \lstick{$q_1$}&\ctrl{1} & \gate{U_2}  & \qw  & \qw\\
    \lstick{$q_2$} & \gate{U_1} & \qw& \gate{U_3} & \qw
    \end{quantikz} $\Rightarrow$ 
    \begin{quantikz}[row sep=2mm]
    \lstick{$q_0$}& \qw & \ctrl{1}\gategroup[2,steps=2,style={dashed,rounded corners, inner xsep=0.2pt},background]{{Optimization}}  &\swap{1} & \qw & \qw\\
    \lstick{$q_1$}&\ctrl{1} & \gate{U_2} & \targX{}& \ctrl{1} & \qw\\
    \lstick{$q_2$} & \gate{U_1} & \qw& \qw &\gate{U_3} & \qw
    \end{quantikz} $\Rightarrow$ 
    \begin{quantikz}[row sep=2mm]
    \lstick{$q_0$}& \qw & \gate{V_1}& \ctrl{1} & \gate{V_2} & \ctrl{1} &\gate{V_3} &\ctrl{1} & \gate{V_4} & \qw & \qw\\
    \lstick{$q_1$}&\ctrl{1} &\gate{V_5} & \targ{} & \gate{V_6} & \targ{} & \gate{V_7}& \targ{}&\gate{V_8}& \ctrl{1} & \qw\\
    \lstick{$q_2$} & \gate{U_1} & \qw & \qw & \qw & \qw & \qw &\qw& \qw &\gate{U_3} & \qw
    \end{quantikz}
    
    \end{adjustbox}
    }
\caption{Two different SWAP gate insertions with the same SWAP gate count but different CNOT gate counts.}
\label{fig:SWAPcost_example}
\end{figure}

In this paper, we make the key observation that the aforementioned compilation flow and the cost functions have several shortcomings. First, the qubit routing step and the circuit optimization step should not be independent. Finding the shortest path with the minimum number of SWAP gates at the routing step may not lead to the optimal design. The reason is that the inserted SWAP gates can be optimized by the subsequent optimizations. When considering the optimization opportunities at the routing step, the SWAP gates should not be treated equally. Some of the SWAP gates may lead to fewer CNOT gates than others. Therefore, rather than the number of inserted SWAP gates, we propose to use the number of inserted CNOT gates as the cost metric since not all of the SWAP gates lead to the same numbers of CNOT gates. We illustrate this observation with an example in Figure~\ref{fig:SWAPcost_example}. Assume that the circuit to be executed consists of pairwise 2-qubit operations, one between qubit 1 and 2, one between qubit 0 and 1, and one between qubit 0 and 2. Also assume that our target device has linear connectivity, i.e., $q_1$ is connected with $q_0$ and $q_2$, but $q_0$ and $q_2$ are not directly connected. As a result, to perform the CNOT gate between $q_0$ and $q_2$, we need to insert SWAP gates and there are two options: either insert a SWAP gate between ($q_0$, $q_1$) or between ($q_1$, $q_2$). If we only consider the SWAP-gate count at the routing step, both routing options have the same SWAP gate cost as one SWAP. Therefore, the compiler may randomly select between these two designs. However, if we consider the subsequent optimizations that would re-synthesize the consecutive two-qubit gates, these two routes actually have different costs in the number of CNOT gates. As shown in Figure~\ref{fig:SWAPcost_example}, the second routing option only needs to insert one CNOT gate, while the first one needs three CNOT gates.

The second shortcoming of the state-of-the-art compilation schemes is that a fixed template is used to decompose the SWAP gates, losing the logic information that the two qubits of the SWAP gate are interchangeable. As a result, fixed SWAP gate decomposition may result in reduced optimization opportunities.
In our approach, we propose an optimization-aware SWAP gate decomposition to overcome this problem. 

We implemented our NASSC in Qiskit v0.28 and compared it with a state-of-the-art scheme, SABRE~\cite{li2019sabre}.  
Our experiments show that the routing overhead compiled with our routing algorithm is reduced by up to $69.30\%$ ($21.30\%$ on average) in the number of CNOT gates and up to $43.50\%$ ($7.61\%$ on average) in the circuit depth compared with SABRE.

Our contributions are summarized as follows:
\begin{itemize}
\item We highlight that qubit routing should not be independent upon the subsequent gate optimizations. 
\item Besides optimization-aware qubit routing, we propose optimization-aware SWAP gate decomposition to facilitate subsequent optimizations.
\item We show that our proposed NASSC algorithm achieves much better results than the prior work. 
\end{itemize}

The remainder of the paper is organized as follows. Section~\ref{sec:background_related_work} introduces the background and the related work. Section~\ref{sec:motivation} presents our observations that motivate optimization-aware qubit routing. Section~\ref{sec:NASSC_structure} discusses the overall compilation process and details our proposed NASSC algorithm. 
Section~\ref{sec:method} describes our compiler implementation and the benchmark set used in our evaluation. Section~\ref{sec:evaluation} presents our experimental results on different hardware topologies. Finally, Section~\ref{sec:conclusions} concludes the paper.
\section{Background and Related Work}
\label{sec:background_related_work}
In this section, we briefly introduce the basic concepts of quantum computing and the structure of the quantum compiler. We also discuss the related works.

\subsection{Quantum Computing}
\label{subsec:quantumcomputing}
Analogous to classical bits, qubits (quantum bits) are the basic unit in quantum computing. A qubit can not only stay in the classical states, $\ket{0}$ and $\ket{1}$, it can also stay in the superposition of these two states. The superposition state is expressed as $\ket{\psi} = a\ket{0} + b\ket{1}$, where $a$ and $b$ are complex numbers and $\left | a \right |^2 + \left | b \right |^2= 1$. An n-qubit quantum system can exist in a superposition of $2^n$ states, which can be represented by a $2^n$ vector of complex values. Besides superposition, entanglement is another unique feature in quantum computing. Qubits can be entangled via two-qubit operations such as CNOT gates. When qubits are entangled, their measurement results are correlated.

A quantum program is a sequence of quantum gates that operate on a number of qubits. An n-qubit quantum gate can be represented by a $2^n\times 2^n$ unitary matrix, $U$. The gate operation can be considered as multiplying the unitary matrix $U$ with the input state $\ket{\psi_0}$. The result is the output state $\ket{\psi_1} = U\ket{\psi_0}$. Some quantum gates in a quantum program may commute, and the compiler optimizes the quantum gates based on commutation analysis. The target quantum hardware may only support a small set of basis gates. For example, the basis gates in the IBM Q system are $id$, $rz$, $sx$, $x$, and $cx$~\cite{IBM_toronto_device}. A quantum gate with a higher-level abstraction such as a Toffoli gate needs to decompose to the basis gates.

\subsection{Quantum Compiler}
\label{subsec:quantumcompiler}
\begin{figure}[htbp]
\centering
    \includegraphics[height = 80mm]{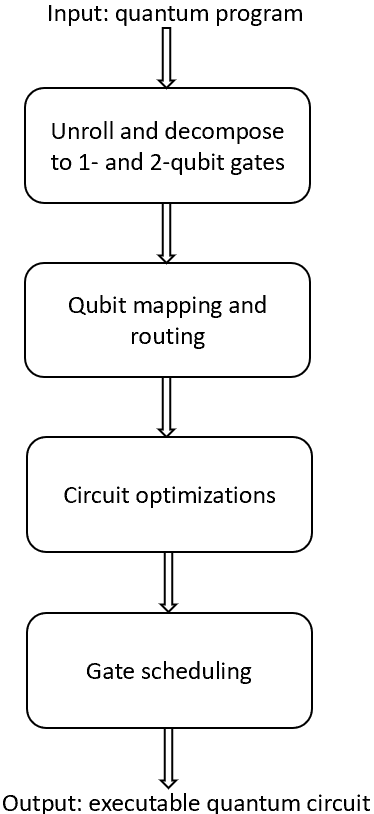}
  \caption{The compilation flow of IBM Qiskit.}
\label{fig:compileflow}
\end{figure}

Quantum compiler plays a critical role in practical quantum computation. A typical compilation pass includes four steps. The first step is decomposing the quantum gates to the basis gates supported by the quantum hardware. Different quantum hardware might support a different set of basis gates. The second step selects logical-to-physical qubit mapping and inserts SWAP gates to route the qubits. Qubit mapping and routing, also known as the qubit allocation problem, has been proven to be NP-hard~\cite{cowtan2019routingnphard}. There has been extensive research using heuristic algorithms or converting the problem to other well-studied problems tackled with classical solvers. The third step performs optimizations to the quantum circuit. We will discuss the common optimizations in Section~\ref{subsec:relatedwork}. The last step schedules the quantum gates to achieve the minimal program runtime and/or highest fidelity. In Figure~\ref{fig:compileflow} we show the simplified step-wise compilation flow of Qiskit. A different compiler may have different compilation flows. For example, the $t\ket{ket}$ compiler~\cite{sivarajah2020tket} has an architecture-independent optimization phase followed by an architecture-dependent phase that prepares the circuit for the target hardware. The architecture-dependent phase includes gate decomposition, qubit mapping and routing, and gate optimizations. Although the compilation flows may differ, the qubit routing and the gate optimizations are independent steps in these compilers. Such independent design may lose optimization opportunities and lead to sub-optimal routing decisions. Based on such findings, we introduce our optimization-aware qubit routing scheme NASSC, which overcomes this potential design limitation. 

\subsection{Related work}
\label{subsec:relatedwork}
First, we review qubit mapping and routing algorithms. The evaluation metric in prior works can be classified into three categories: circuit size (i.e., number of extra gates inserted at the routing step; which is often equivalent to the number of extra SWAP gates), circuit depth (i.e., the number of layers in the final circuit), and the circuit error rate.

Zulehner et al.~\cite{zulehner2018efficient} proposed an approach to partition quantum circuits into layers. Each layer contains gates that can be executed in parallel. Then for each layer, SWAP gates are inserted to find a hardware compliant mapping. The A* algorithm is adopted to search for the path with the lowest cost, where the cost is the number of elementary operations. While their cost function considers the basis gates, their SWAP gates always follow the same decomposition and have the identical cost of seven elementary gates (three CNOTs and four Hadamard operations for monodirectional links). Li et al.~\cite{li2019sabre} proposed a SWAP-based bidirectional heuristic search method named SABRE. SABRE first computes a distance matrix of the target hardware. The best route is selected using a heuristic cost function based on the distance matrix. The lookahead cost function considers not only the front layer but also the subsequent layers. They also leverage intrinsic reversibility to enable global optimization. Inspired by SABRE, HA is a hardware-aware heuristic proposed by Niu et al.~\cite{niu2020hardware}. HA improved the fidelity and reduced the number of additional gates by introducing a new distance matrix based on hardware connectivity and calibration data. The cost function estimates the success rate of the inserted gates. Itoko et al.~\cite{itoko2020mapping_commutation} takes advantage of the commutation rules to find the gates that commute in the original circuit. Such an approach can explore more routing candidates than the fixed layer approach. However, the routing algorithm is not optimization-aware as the CNOT gate count for the SWAP gate decomposition always stays the same. Besides these, there are prior works that use the number of SWAP gates as the cost function and propose different approaches to find the best route~\cite{zhu2020dynamicswapcount1,zhou2020monteswapcount2,childs2019circuitswapcount3,wille2019minimumSWAP}. There are also works that incorporate the error rates in their cost functions~\cite{ash2019qurenoise1, murali2019noise,patel2020ureqanoise3}. Some other prior works also take the circuit depth into consideration~\cite{zhang2020slackqdepth1,zhang2020depth2}. In summary, none of these prior works on qubit routing is optimization aware. 

Next, we review various quantum circuit optimizations. Peephole optimization~\cite{mckeeman1965peephole, liu2021relaxedpeephole} is widely used in quantum compilers. The peephole optimization identifies subcircuits in specific patterns and substitutes them with equivalent circuits that have lower cost. The Qiskit transpiler~\cite{Qiskit} contains the \texttt{Optimize1qGates} optimization pass, which identifies the pattern of consecutive single-qubit gates and substitutes them altogether with a single-qubit gate. The transpiler also contains optimization passes \texttt{Collect2qBlocks} and \texttt{UnitarySynthesis} to identify the two-qubit blocks and re-synthesize them. The $t\ket{ket}$~\cite{sivarajah2020tket} compiler identifies long sequences of single-/two-qubit gates and re-synthesizes them with Euler and KAK decomposition~\cite{kraus2001kak1}. Similar optimizations can also be found in the Cirq~\cite{cirq} compiler. These optimizations can be considered as in the category of peephole optimization. Commutation analysis has been utilized for gate optimization~\cite{miller2004reversiblecommute} and qubit mapping~\cite{alam2020qaoamapping,itoko2020mapping_commutation}. The compiler can identify more templates by reordering the quantum gates. In Qiskit, the \texttt{CommutationAnalysis} pass finds the commutation relations between the quantum gates, and groups the gates in a set of gates that commute. After commutation analysis, the gates are optimized with gate cancellation~\cite{maslov2008cancellation}. Qiskit also has optimization passes including noise adaption~\cite{murali2019noise}, crosstalk mitigation~\cite{murali2020softwarecrosstalk}, and scheduling optimization~\cite{shi2019optimized_scheduling}.

\section{Motivation}
\label{sec:motivation}
In this section, we motivate the need for optimization-aware qubit routing. We look closely at the Qiskit compiler framework and study the optimization passes that may change the cost of the SWAP gates.

Gate optimizations can remarkably reduce the count of the basis gates inserted at the qubit routing step. As we studied the quantum circuit generated by Qiskit, we found that a large proportion of the inserted SWAP gates are actually modified by the subsequent optimizations, the re-synthesis of two-qubit blocks and gate cancellation, in particular. For example, when a 10-qubit Grover benchmark is mapped to a backend with $4\times 4$ 2D-grid connectivity, $20.7\%$ of the SWAP gates are optimized by the two-qubit block re-synthesis while $40.3\%$ of them are optimized by gate cancellation. Such observation indicates that even the compiler selects the path with the fewest number of SWAP insertions at the routing step, the final cost of the selected route may not be optimal. 

\begin{figure}[htbp]
        \centering
    \begin{quantikz}
    \lstick{}&\gate{V_1}\gategroup[2,steps=4,style={dashed,rounded corners, inner xsep=0.2pt},background]{{two-qubit block}} & \ctrl{1} & \gate{V_2} & \swap{1} & \qw\\
    \lstick{}&\gate{V_3} & \targ{}  & \gate{V_4}  & \targX{} & \qw
    \end{quantikz} $\Rightarrow$ 
    \begin{quantikz}
    \lstick{}&\gate{U_1} & \ctrl{1} & \gate{U_2} & \ctrl{1} & \gate{U_3} & \ctrl{1} & \gate{U_4} & \qw\\
    \lstick{}&\gate{U_5} & \targ{}  & \gate{U_6}  & \targ{} & \gate{U_7} & \targ{}& \gate{U_8} &\qw
    \end{quantikz}
    
     \caption{The re-synthesis of two-qubit block and the universal two-qubit gate decomposition reduce the cost of the SWAP gate. }
\label{fig:2qblock}
\end{figure}
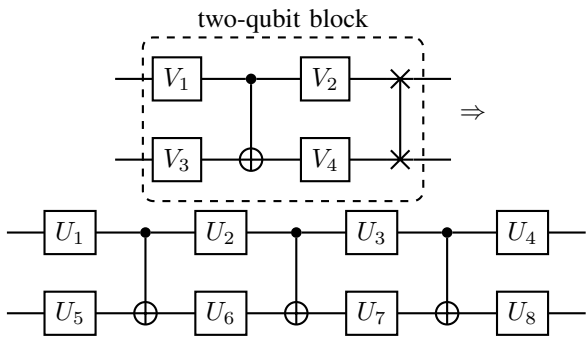

The first optimization that optimizes the SWAP gate is the re-synthesis of two-qubit blocks. A two-qubit block~\cite{Qiskit} is a sequence of uninterrupted two-qubit gates. Since the two-qubit block operator $U \in SU(4)$, it can be generated with a  two-qubit gate with three CNOTs~\cite{vidal2004universal2qblock}. The compiler calculates the matrix representation of the two-qubit block and uses the KAK decomposition~\cite{kraus2001kak1} to generate a subcircuit with up to three CNOT gates. Figure~\ref{fig:2qblock} shows an example of this optimization. In this example, after re-synthesis, the cost of implementing the SWAP gate is two CNOT gates and several single-qubit gates. In some extreme cases, when the gate sequence already contains at least three CNOTs before the SWAP insertion, the SWAP gate can be inserted for free. In other words, some SWAP gates can be inserted at no cost!

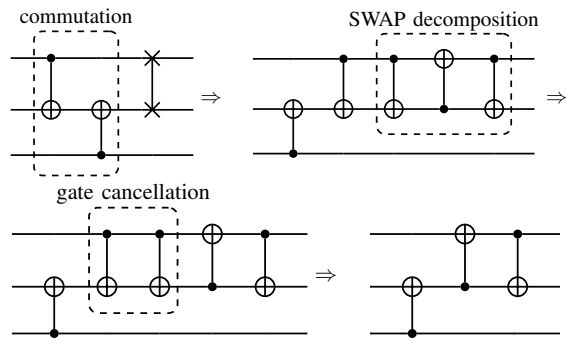
\begin{figure}[htbp]
        \centering
   \begin{adjustbox}{width=0.9\linewidth}
    \begin{quantikz}
    \lstick{} & \ctrl{1}\gategroup[3,steps=2,style={dashed,rounded corners, inner xsep=0.2pt},background]{{commutation}} & \qw & \swap{1} & \qw\\
    \lstick{} & \targ{}  & \targ{}  & \targX{} & \qw\\
    \lstick{} & \qw  & \ctrl{-1}  & \qw & \qw
    \end{quantikz} $\Rightarrow$ 
    \begin{quantikz}
    \lstick{} & \qw & \ctrl{1} & \ctrl{1}\gategroup[2,steps=3,style={dashed,rounded corners, inner xsep=0.2pt},background]{{SWAP decomposition}} & \targ{} & \ctrl{1} &\qw\\
    \lstick{} & \targ{}  & \targ{}  & \targ{}& \ctrl{-1} & \targ{}&\qw\\
    \lstick{} & \ctrl{-1}  & \qw  & \qw & \qw & \qw &\qw
    \end{quantikz}
    $\Rightarrow$
    \end{adjustbox}
   \begin{adjustbox}{width=0.9\linewidth} 
        \begin{quantikz}
    \lstick{} & \qw & \ctrl{1}\gategroup[2,steps=2,style={dashed,rounded corners, inner xsep=0.2pt},background]{{gate cancellation}} & \ctrl{1} &\targ{} & \ctrl{1} & \qw\\
    \lstick{} & \targ{}  & \targ{}  & \targ{} &\ctrl{-1}  & \targ{}& \qw\\
    \lstick{} & \ctrl{-1}  & \qw  & \qw & \qw & \qw & \qw
    \end{quantikz}$\Rightarrow$ 
    \begin{quantikz}
    \lstick{} & \qw & \targ{} & \ctrl{1} & \qw\\
    \lstick{} & \targ{}  & \ctrl{-1} & \targ{} & \qw\\
    \lstick{} & \ctrl{-1}  & \qw  & \qw & \qw
    \end{quantikz}
    \end{adjustbox}
     \caption{SWAP gate optimization with gate commutation and cancellation.}
\label{fig:gatecommutation}
\end{figure}

The second optimization that may affect the cost of SWAP gates is gate cancellation. The compiler will search for the potential cancellable quantum gates based on commutation analysis. In Qiskit, the \texttt{CommutativeCancellation} pass cancels the self-inverse gates through commutation relations. The following self-inverse gates are considered: $H$, $X$, $Y$, $Z$, $CX$, $CY$, and $CZ$. We show an example in Figure~\ref{fig:gatecommutation} to illustrate this optimization. In the first circuit, the first two CNOT gates commute since they share the same target qubit~\cite{itoko2020mapping_commutation}. If we switch the order of the first two CNOT gates, an inserted CNOT gate can be canceled with the second CNOT gate in the original circuit. Therefore, the number of CNOT gates required by the SWAP is no longer three. In this example, the SWAP gate will only introduce one extra CNOT gate. In other words, the cost of a SWAP is not a fixed value and is dependent on the subsequent optimizations.

Both of the optimizations can reduce the CNOT gate count required by SWAP gates. This finding leads to the design of our optimization-aware routing algorithm. 
As a matter of fact, the above-mentioned optimizations may not identify all the potential optimization opportunities for SWAP gates. In Section~\ref{subsec:commutation}, we will discuss the SWAP-related gate optimization and our new optimization-aware SWAP gate decomposition. 

\section{NASSC}
\label{sec:NASSC_structure}
In this section, we describe our proposed NASSC approach. In Section~\ref{subsec:overview}, we show an overview of the NASSC algorithm and its integration with Qiskit. In Section~\ref{subsec:heuristic}, we present our search heuristic. In Section~\ref{subsec:costfunction} we discuss our cost function. In Section~\ref{subsec:2qresynthesis} and Section~\ref{subsec:commutation} we discuss the two optimizations that can impact the cost of SWAP gates. We discuss the integration of multiple optimizations in Section~\ref{subsec:strategy}.

\begin{figure}[htbp]
\centering
    \includegraphics[width = \linewidth]{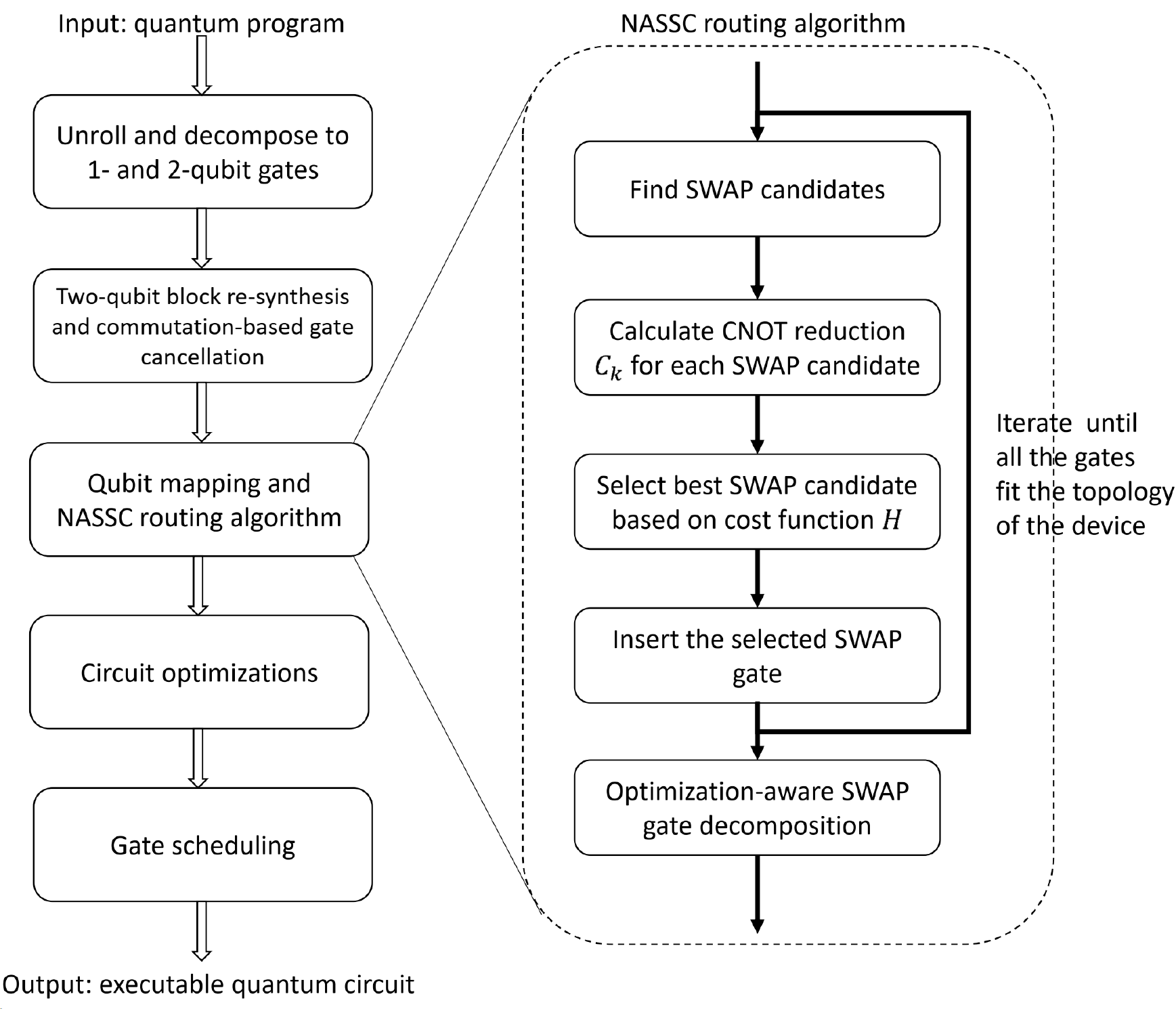}
  \caption{The compilation flow of NASSC integrated with Qiskit}
\label{fig:nassc_compileflow}
\end{figure}

\subsection{Overview}
\label{subsec:overview}
An overview of NASSC and its integration with Qiskit is shown in Figure~\ref{fig:nassc_compileflow}. Our routing algorithm considers the two-qubit block re-synthesis and commutation-based gate cancellation at the routing step. In order to collect the optimization information and shorten the transpilation time, we move the corresponding optimizations before qubit mapping and routing. We use the same qubit mapping algorithm as SABRE, which includes the random front layer initialization and the reverse traversal initial mapping update~\cite{li2019sabre}. After the initial mapping, for each two-qubit gate that does not fit on the device, the NASSC routing algorithm finds the SWAP candidates that might move the logical qubits closer. For each SWAP candidate, the algorithm identifies the potential optimizations and calculates CNOT gate count reduction $C_k$ with each optimization. Some of the optimizations might require a special SWAP decomposition. The compiler will flag the SWAP candidates with the special decomposition requirement. Then the algorithm calculates the cost function for each SWAP candidate and inserts the SWAP gate with the smallest cost. The algorithm will iteratively insert SWAP gates until all the two-qubit gates fit the topology of the device. After inserting all the SWAP gates, the last step of the NASSC algorithm is performing optimization-aware SWAP gate decomposition based on the flags. After the routing step, the compiler will perform the rest of the circuit optimizations and schedule the gates.

\subsection{Optimization-Aware Qubit Routing}
\label{subsec:heuristic}
In our routing algorithm, we first reformulate the logical quantum circuit representation to the Directed-Acyclic-Graph (DAG) format. The DAG is constructed such that a DAG node represents a quantum gate, and the directed edge (i,j) between node i and node j represents the dependency from node i to node j. In other words, gate i should be executed before gate j.

\begin{figure*}[t]
        \centering
    \subfloat[Connectivity graph\label{subfig:layered_connectivity}]{
    \includegraphics[width = 0.065\linewidth, valign=c]{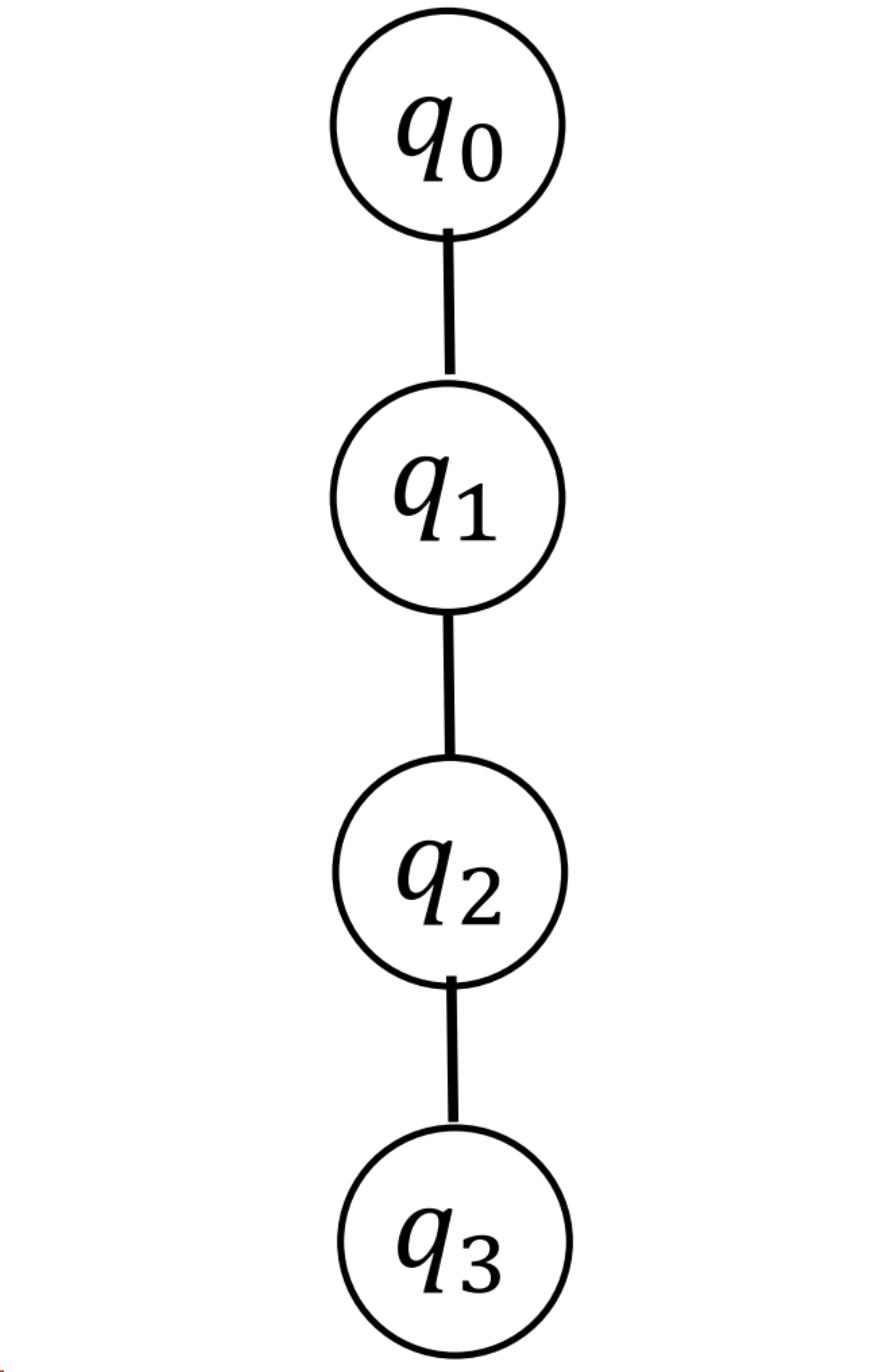}
    }
    \subfloat[Original circuit\label{subfig:layered_original}]{  
    \begin{adjustbox}{width=0.2\linewidth}
    \begin{quantikz}
    \lstick{$q_0$}& \qw\gategroup[4,steps=2,style={dashed,rounded corners, inner xsep=0.2pt},background]{{\footnotesize Resolved}}&\ctrl{1} & \ctrl{2}\gategroup[3, steps=1,style={dashed,rounded corners, inner xsep=0.2pt},background]{{\footnotesize Front}} &\qw\gategroup[3, steps=2,style={dashed,rounded corners, inner xsep=0.2pt},background]{{\footnotesize Extended}} & \ctrl{1} & \qw\\
    \lstick{$q_1$}&\targ{} &\gate{Rx} & \qw &\ctrl{1} & \targ{} & \qw\\
    \lstick{$q_2$}& \ctrl{-1} & \ctrl{1} & \targ{} &\gate{Rx}&\qw &\qw\\
    \lstick{$q_3$}&\qw &  \targ{} & \qw &\qw & \qw & \qw
    \end{quantikz}
    \end{adjustbox}
    }
    \subfloat[SWAP insertion option 1\label{subfig:layered_option1}]{  
    \begin{adjustbox}{width=0.2\linewidth}
    \begin{quantikz}
    \lstick{$q_0$}& \qw\gategroup[4,steps=2,style={dashed,rounded corners, inner xsep=0.2pt},background]{{\footnotesize Resolved}}&\ctrl{1} & \swap{1} & \qw\gategroup[3, steps=1,style={dashed,rounded corners, inner xsep=0.2pt},background]{{\footnotesize Front}} &\ctrl{2}\gategroup[3, steps=2,style={dashed,rounded corners, inner xsep=0.2pt},background]{{\footnotesize Extended}} & \targ{} & \qw\\
    \lstick{$q_1$}&\targ{} &\gate{Rx} &\targX{} & \ctrl{1} &\qw & \ctrl{-1} & \qw\\
    \lstick{$q_2$}& \ctrl{-1} & \ctrl{1} & \qw &\targ{} &\gate{Rx}&\qw &\qw\\
    \lstick{$q_3$}&\qw &  \targ{} & \qw &\qw & \qw & \qw & \qw
    \end{quantikz}
    \end{adjustbox}
    }
    \subfloat[SWAP insertion option 2\label{subfig:layered_option2}]{  
    \begin{adjustbox}{width=0.2\linewidth}
    \begin{quantikz}
    \lstick{$q_0$}& \qw\gategroup[4,steps=2,style={dashed,rounded corners, inner xsep=0.2pt},background]{{\footnotesize Resolved}}&\ctrl{1} & \qw & \ctrl{1}\gategroup[3, steps=1,style={dashed,rounded corners, inner xsep=0.2pt},background]{{\footnotesize Front}} &\qw\gategroup[3, steps=2,style={dashed,rounded corners, inner xsep=0.2pt},background]{{\footnotesize Extended}} & \ctrl{2} & \qw\\
    \lstick{$q_1$}&\targ{} &\gate{Rx} &\swap{1} & \targ{} &\gate{Rx} & \qw & \qw\\
    \lstick{$q_2$}& \ctrl{-1} & \ctrl{1} & \targX{} &\qw &\ctrl{-1}&\targ{} &\qw\\
    \lstick{$q_3$}&\qw &  \targ{} & \qw{} &\qw & \qw & \qw & \qw
    \end{quantikz}
    \end{adjustbox}
    }
    \subfloat[Circuit after SWAP insertion\label{subfig:layered_final}]{  
    \begin{adjustbox}{width=0.2\linewidth}
    \begin{quantikz}
    \lstick{$q_0$}& \qw\gategroup[4,steps=4,style={dashed,rounded corners, inner xsep=0.2pt},background]{{\footnotesize Resolved}}&\ctrl{1} & \qw & \ctrl{1} &\qw\gategroup[3, steps=1,style={dashed,rounded corners, inner xsep=0.2pt},background]{{\footnotesize Front}} & \ctrl{2}\gategroup[3, steps=1,style={dashed,rounded corners, inner xsep=0.2pt},background]{{\footnotesize Extended}} & \qw\\
    \lstick{$q_1$}&\targ{} &\gate{Rx} &\swap{1} & \targ{} &\gate{Rx} & \qw & \qw\\
    \lstick{$q_2$}& \ctrl{-1} & \ctrl{1} & \targX{} &\qw &\ctrl{-1}&\targ{} &\qw\\
    \lstick{$q_3$}&\qw &  \targ{} & \qw{} &\qw & \qw & \qw & \qw
    \end{quantikz}
    \end{adjustbox}
    }
     \caption{The layered search heuristic.}
\label{fig:layer_heuristic}
\end{figure*}

Then, the quantum gates in the circuit are divided into three groups: the resolved gates, the executable gates, and the to-be-executable gates. The resolved gates are the ones that have already been mapped by the algorithm, including the inserted SWAP gates. The resolved gates form the resolved layer. The executable gates are the gates that do not have un-executed predecessors in the DAG. The executable form the front layer. The to-be-executable gates are the rest of the gates. Some of the closest successors of the gates in the front layer form the extended layer. The extended layer is set for lookahead analysis. The abstraction of different layers helps to define the search heuristic. And an illustration of these layers is shown in Figure~\ref{fig:layer_heuristic}.

Next, we perform layered search to determine qubit routing. In our algorithm, we use the random front layer initialization and the reverse traversal initial mapping update, the same as SABRE~\cite{li2019sabre}. For the gates in the front layer, the algorithm will remove the gates that are directly executable with the current mapping and add them to the resolved layer. The remaining gates in the front layer, if there are any, would require SWAP gates insertion.  In the example shown in Figure~\ref{fig:layer_heuristic}, assume that the circuit is to be executed on a hardware backend with linear-nearest neighbor connectivity. Then, the CNOT gate in the front layer in Figure~\ref{fig:layer_heuristic}a is not directly executable since physical qubits $q_0$ and $q_2$ are not connected. In this case, the compiler needs to find SWAP candidates. For each logical qubit in the front layer, the compiler finds its current physical qubit and its adjacent physical qubits to construct potential routing candidates. Every SWAP corresponding to one of the physical couplings is considered a candidate SWAP. For the example in Figure~\ref{fig:layer_heuristic}, $q_0$ is the physical qubit that correlates to the logical qubit in the front layer, and $q_1$ is its adjacent qubit. Hence the SWAP between $q_0$ and $q_1$ is a candidate SWAP as shown in Figure~\ref{subfig:layered_option1}. Similarly we can have another candidate SWAP between $q_1$ and $q_2$ as shown in Figure~\ref{subfig:layered_option2}. All of the candidate SWAPs form a SWAP candidate set $T$. The compiler will calculate the cost of each SWAP candidate in $T$ based on an optimization-aware cost function $H$. For a SWAP candidate, the compiler will analyze its predecessor and successors to estimate the cost. In Figure~\ref{subfig:layered_option1}, the SWAP gate between $q_0$ and $q_1$ can be resynthesized with the controlled-Rx gate. Therefore, it will introduce two CNOT gates. The SWAP gate between $q_1$ and $q_2$ in Figure~\ref{subfig:layered_option2} will be optimized with gate cancellation and introduce one CNOT gate. The reason is that the CNOT between $q_1$ and $q_2$ commutes with the controlled $R_x$ gate between $q_0$ and $q_1$. As a result, the CNOT can cancel one of the CNOTs decomposed from the SWAP gate.
As shown in Figure~\ref{subfig:layered_final}, the SWAP candidate with the minimum cost is selected. For a gate that requires multiple hops, the compiler iteratively inserts SWAP gates, one hop a time until the gate becomes implementable on the target device. Then, the above process repeats until all of the gates in the front layer are resolved. Once the front layer is empty, the compiler will select the executable gates that form a new front layer, as shown in Figure~\ref{subfig:layered_final}. The compiler finishes routing when the front layer and the extended layer are both empty. During SWAP gate insertion, the compiler will maintain the information such as the total number of SWAPs and collect the information relating to the SWAP candidate set $T$.

\subsection{Optimization-Aware Cost Function}
\label{subsec:costfunction}
The cost function is crucial for qubit routing as it is used to select the optimal route from the candidate routing sets. The basic cost function in NASSC is defined as:
\begin{equation} 
    \resizebox{0.9\linewidth}{!}{%
$\begin{aligned}
    H_{basic} = \sum_{gate\in F}3\times D[g.i][g.j] - \sum_{k\in opt} b_{k} \times C_{k}
\end{aligned}$
}
\end{equation}

The cost function is calculated for every SWAP candidate: for a two-qubit gate, the candidates include all the couplings/connections from either of the two physical qubits of the gate. The first part of the equation calculates the number of CNOT gates without considering optimization, which is why the constant 3 is used as one SWAP can be implemented with 3 CNOT gates. The second part of the equation estimates the CNOT reduction for the SWAP gates. In the equation, $D$ is the distance matrix, which records the distance between different physical qubits. $D[g.i][g.j]$ represents the distance between the input qubits of gate $g$.
$g.i$ and $g.j$ are the physical qubits of the gate $g$ %before and 
after SWAP insertions (the cost before SWAP insertion is the same for all the SWAP candidates, thereby not being included in the cost function).
$F$ represents the front layer. $b_k$ and $C_k$ are introduced to reflect the cost reduction resulting from the subsequent gate optimizations. $b_k\in\{0,1\}$ is the binary value for the $k$th optimization in the set of gate optimizations $opt$. Since different benchmarks might favor different optimizations, we use binary value $b_k$ to enable/disable the $k$th optimization. The discussion can be found in Section~\ref{subsec:strategy}. In our experimental results, the binary values $b_k$ are set to 1 to enable all the optimizations. $C_k$ is the estimated CNOT gate count reduction with the $k$th optimization. 
In this work, we consider the following optimizations in $opt$ as they are related to SWAP gates: two-qubit block re-synthesis ($C_{2q}$), and two commutation-related optimizations ($C_{commute1}, C_{commute_2}$).

Although our basic cost function in eq. 1 considers both the routing distance and the impact from subsequent optimizations, it might lead to a local optimal result. The reason is that finding the best route for the gates in the front layer might hinder the routing of the gates in the extended layer. Therefore, the extended layer cost should be added to the cost function to increase the lookahead capability of the algorithm, and the resulting cost function is defined as: 
 
\begin{equation} 
\resizebox{0.9\linewidth}{!}{%
$\begin{aligned}
    H = H_{basic} + H_{lookahead} = \frac{1}{|F|}(3\times \sum_{gate\in F}D[g.i][g.j] - \\
    \sum_{k\in opt} b_{k} \times C_{k})
    + \frac{W}{|E|}\sum_{gate\in E}D[g.i][g.j]
\end{aligned}$
}
\end{equation}

 In the equation, $E$ is the extended layer and we can adjust its size to account for different lookahead windows. The cost of the extended layer is the routing distances based on the distance matrix. The impacts of the front and the extended layer are normalized by their sizes. A weight parameter $W$ is introduced to prioritize the impact of SWAPs in the front layer. 
 
 Since we move the two-qubit block re-synthesis and commutation analysis before the qubit routing pass, for every SWAP candidate, calculating corresponding $C_k$ only requires constant overhead. For the two-qubit block re-synthesis, calculating $C_{2q}$ requires the re-synthesis of the two-qubit block with a SWAP gate. Considering that pre-routing optimizations have re-synthesized the qubit block into a concise form, re-synthesizing that block and calculating $C_{2q}$ have time complexity of $O(1)$. For the commutation-related optimizations, CNOT gate count reductions $C_{commute1}$, and $C_{commute2}$ are based on the commutation set information from the commutation analysis pass. Calculating $C_{commute1}$, and $C_{commute2}$ have complexity $O(1)$. The detailed calculation of $C_k$ can be found in Section~\ref{subsec:2qresynthesis} and ~\ref{subsec:commutation}. The time complexity of the heuristic cost function is $O(N)$ since there are at most $O(N)$ gates in the front layer. Here $N$ is the number of physical qubits. The extended layer size is not considered since it is not very large and is set to a fixed number in our experiments, i.e., $O(1)$. Strictly speaking, the cost of the extended layer should also consider the impact of optimizations as well. We ignore such impact in the extended layer for two reasons. First, the impact from lookahead/the extended layer is relatively small compared to the cost of the front layer. Second, since the circuit structure in the extended layer has not been fixed, the calculation for the optimization impact on SWAPs in the extended layer would be imprecise. 

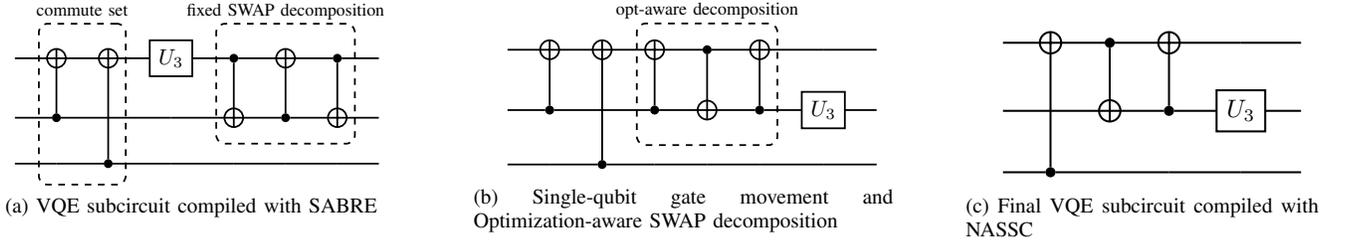
\begin{figure*}
    \subfloat[VQE subcircuit compiled with SABRE\label{subfig:vqe_original}]{  
    \begin{adjustbox}{width=0.3\linewidth}
    \begin{quantikz}
    \lstick{}& \targ{}\gategroup[3,steps=2,style={dashed,rounded corners, inner xsep=0.2pt},background]{{\footnotesize commute set}}& \targ{} & \gate{U_3}& \ctrl{1}\gategroup[2,steps=3,style={dashed,rounded corners, inner xsep=0.2pt},background]{{\footnotesize fixed SWAP decomposition}} &\targ{} & \ctrl{1}& \qw\\
    \lstick{}&\ctrl{-1} &\qw &\qw &\targ{} & \ctrl{-1} & \targ{} & \qw\\
    \lstick{}& \qw & \ctrl{-2} & \qw &\qw &\qw &\qw &\qw
    \end{quantikz}
    \end{adjustbox}
    }\hfill
    \subfloat[Single-qubit gate movement and Optimization-aware SWAP decomposition\label{subfig:vqe_intermediate}]{  
    \begin{adjustbox}{width=0.3\linewidth}
    \begin{quantikz}
    \lstick{}& \targ{}& \targ{}& \targ{}\gategroup[2,steps=3,style={dashed,rounded corners, inner xsep=0.2pt},background]{{\footnotesize opt-aware decomposition}} & \ctrl{1} &\targ{}& \qw & \qw\\
    \lstick{}&\ctrl{-1} &\qw & \ctrl{-1} &\targ{} & \ctrl{-1} & \gate{U_3} & \qw\\
    \lstick{}& \qw & \ctrl{-2}  &\qw &\qw &\qw& \qw &\qw
    \end{quantikz}
    \end{adjustbox}
    }\hfill
    \subfloat[Final VQE subcircuit compiled with NASSC\label{subfig:vqe)final}]{  
    \begin{adjustbox}{width=0.25\linewidth}
    \begin{quantikz}
    \lstick{}& \targ{} & \ctrl{1} &\targ{}& \qw & \qw\\
    \lstick{} &\qw &\targ{} & \ctrl{-1} & \gate{U_3} & \qw\\
    \lstick{} & \ctrl{-2}  &\qw &\qw& \qw &\qw
    \end{quantikz}
    \end{adjustbox}
    }
     \caption{Case 1: Gate cancellation between a CNOT and a SWAP gate}
\label{fig:vqecommute}
\end{figure*}

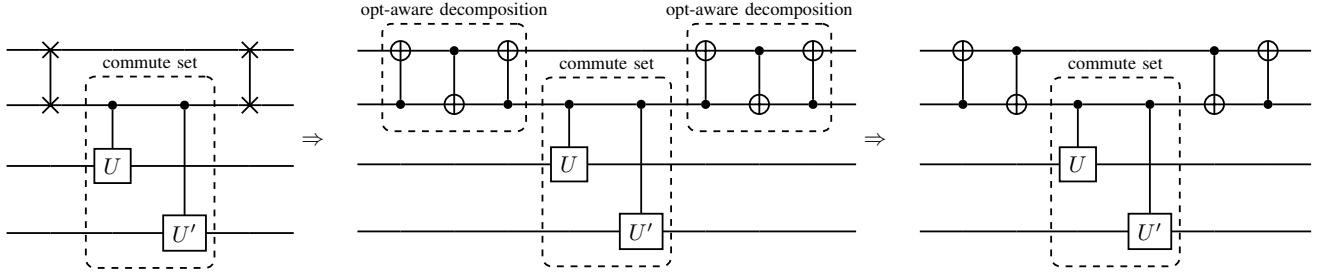
\begin{figure*}[htbp]
        \centering
    \begin{adjustbox}{width=\linewidth}
    \begin{quantikz}
    \lstick{}& \swap{1} & \qw  & 
\qw &\swap{1}& \qw\\
    \lstick{}&\targX{} &\ctrl{1}\gategroup[3,steps=2,style={dashed,rounded corners, inner xsep=0.2pt},background]{{\footnotesize commute set}} & \ctrl{2} & \targX{} & \qw\\
    \lstick{}& \qw & \gate{U}&\qw &\qw &\qw\\
    \lstick{}& \qw & \qw &\gate{U'} &\qw &\qw
    \end{quantikz}$\Rightarrow$
    \begin{quantikz}
    \lstick{}& \targ{}\gategroup[2,steps=3,style={dashed,rounded corners, inner xsep=0.2pt},background]{{\footnotesize opt-aware decomposition}} &\ctrl{1} &\targ{} & \qw  & 
\qw &\targ{}\gategroup[2,steps=3,style={dashed,rounded corners, inner xsep=0.2pt},background]{{\footnotesize opt-aware decomposition}} &\ctrl{1} & \targ{}& \qw\\
    \lstick{}&\ctrl{-1} & \targ{} &\ctrl{-1} &\ctrl{1}\gategroup[3,steps=2,style={dashed,rounded corners, inner xsep=0.2pt},background]{{\footnotesize commute set}} & \ctrl{2} &\ctrl{-1} & \targ{} & \ctrl{-1} & \qw\\
    \lstick{}&\qw & \qw &\qw& \gate{U}&\qw &\qw &\qw & \qw& \qw\\
    \lstick{}& \qw & \qw & \qw &\qw &\gate{U'} &\qw &\qw & \qw & \qw
    \end{quantikz}$\Rightarrow$
    \begin{quantikz}
    \lstick{}& \targ{} &\ctrl{1} & \qw  & 
\qw &\ctrl{1} & \targ{}& \qw\\
    \lstick{}&\ctrl{-1} & \targ{} &\ctrl{1}\gategroup[3,steps=2,style={dashed,rounded corners, inner xsep=0.2pt},background]{{\footnotesize commute set}} & \ctrl{2} & \targ{} & \ctrl{-1} & \qw\\
    \lstick{}& \qw &\qw& \gate{U}&\qw &\qw &\qw & \qw\\
    \lstick{}& \qw & \qw &\qw &\gate{U'} &\qw &\qw & \qw
    \end{quantikz}
    \end{adjustbox}
     \caption{Case 2: CNOTs cancellation across two SWAP gates}
\label{fig:swapcancel}
\end{figure*}

\subsection{Two-qubit Block Re-synthesis}
\label{subsec:2qresynthesis}
As aforementioned, two-qubit block re-synthesis is an optimization that may affect the cost of SWAP gates. In the cost function, the value $C_{2q}$ is used to represent the impact of the CNOT gate reduction from two-qubit block re-synthesis. $C_{2q}$ can be one of the four possible values 0, 1, 2, and 3 depending on how many CNOT gates can be removed. The maximum value is 3 as a SWAP needs at most 3 CNOT gates. Our compiler implementation uses the following way to calculate the number of CNOT reduction. 
In Qiskit, the two-qubit block re-synthesis pass traverses the DAG to find the gates in the same two-qubit block. We move the two-qubit block re-synthesis pass before our qubit routing step for $C_{2q}$ estimation. After the pass, the gates that belong to the same two-qubit block will have the same $block\_id$. For each SWAP in the SWAP candidate set $T$, the compiler checks the $block\_id$ of its predecessors. If they share the same $block\_id$, it means that they are in the same two-qubit block. Then the SWAP candidate can also be included in the same block. The compiler invokes the two-qubit block re-synthesis function to calculate the CNOT gate count difference before and after the SWAP gates insertion. This difference would be the reduced cost $C_{2q}$ for that particular SWAP. Note that since the two-qubit blocks have been re-synthesized into a fixed template before routing, calculating $C_{2q}$ only requires constant overhead $O(1)$.

\subsection{Commutation-Based Optimizations}
\label{subsec:commutation}
Commutation-based optimization can effectively reduce the circuit complexity by exploring the order of gates to find opportunities for gate cancellation. However, the effectiveness of this optimization can be significantly affected by qubit routing and SWAP decomposition. 
For example, Figure~\ref{subfig:vqe_original} shows a snip of the 10-qubit VQE circuit compiled with SABRE. The routed VQE circuit can not be optimized by the subsequent \texttt{CommutativeCancellation} pass, which is the commutation-based optimization implemented in Qiskit. There are two reasons. First, the single-qubit $U_3$ gate does not commute with the first two CNOT gates. As a result, it blocks the commutation analysis. Second, the SWAP gate has already been decomposed into three CNOT gates before the optimization step. The compiler would not be able to find an opportunity for gate cancellation even if the $U_3$ gate is not there. On the other hand, if we retain the semantics of the SWAP operation, we can see that the single-qubit gate $U_3$ can be commuted with the SWAP gate by moving $U_3$ to the swapped qubit. Furthermore, if the compiler is aware that one of the qubits to be swapped is used as the control qubit of a CNOT, it can choose the decomposition such that this qubit is the control bit of the first CNOT gate in the decomposition, as shown in Figure~\ref{subfig:vqe_intermediate}. Then, the commutation-based optimization can discover the gate cancelling opportunities, which would result in a much simpler circuit as shown in Figure~\ref{subfig:vqe)final}. 

Motivated from the example in Figure~\ref{fig:vqecommute}, we propose optimization-aware SWAP gate decomposition at the qubit routing step to facilitate subsequent commutation-based optimizations. There are two possible decompositions for a SWAP gate to be decomposed into three CNOT gates depending on which qubit is used as the control qubit of the first CNOT. For every SWAP candidate in the qubit routing step, the compiler will look for potential commutation-based optimization opportunities. If the first CNOT gate in the SWAP can potentially be cancelled with a CNOT gate in the circuit, the SWAP gate should be decomposed according to the control and the target qubit of the CNOT in the circuit. Also, as shown in Figure~\ref{fig:swapcancel}, if the first and the last CNOT gate in two SWAP gates can be potentially cancelled through commutation, both SWAP gates should be decomposed according to the intermediate commute set. 

Next, we derive the cost reduction from commutation-based optimizations, i.e., $C_k$ in the cost function. We consider commutation-based optimizations in two scenarios. The first is that a decomposed SWAP gate can be optimized with CNOT gates in the circuit, similar to the example shown in Figure~\ref{fig:vqecommute}. The second is that a decomposed SWAP gate can be optimized with previously inserted SWAP gates. An example is shown in Figure~\ref{fig:swapcancel}, where a set of commutable gates is sandwiched by two SWAP gates.   
Once the gates form this sandwich-like structure, both SWAP gates can reduce their CNOT cost by one, as shown in the figure. Since these two scenarios happen frequently and have different circuit structures, we treat them as individual optimizations and derive different cost reductions in the cost function. 

\begin{figure*}
    \centering
    \subfloat[\texttt{ibmq\_montreal} \label{fig:CNOT_reduction_compare_montreal}]{  \includegraphics[width = 0.3\linewidth]{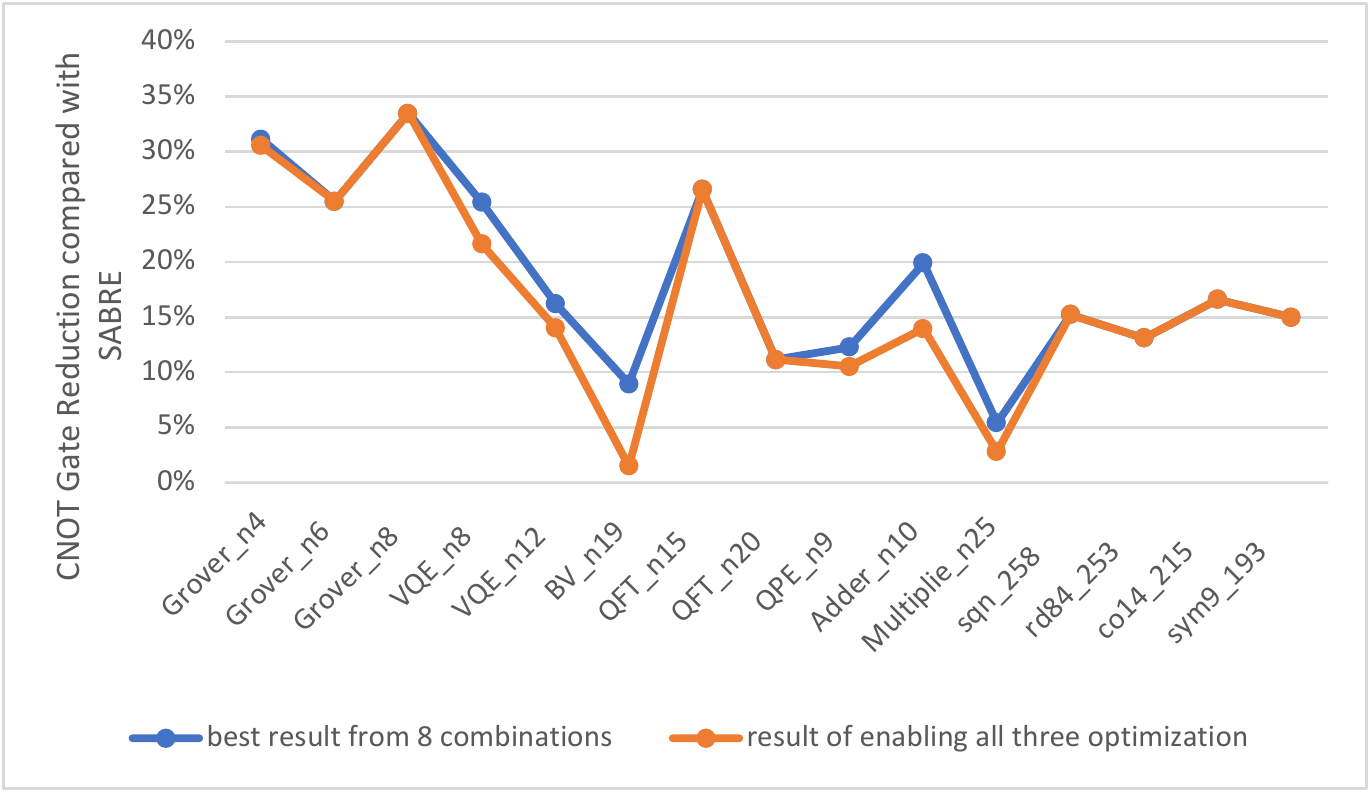}
    }\hfill
    \subfloat[\texttt{25-qubit linear nearest neighbour\label{fig:CNOT_reduction_compare_linear}}]{      \includegraphics[width=0.3\linewidth]{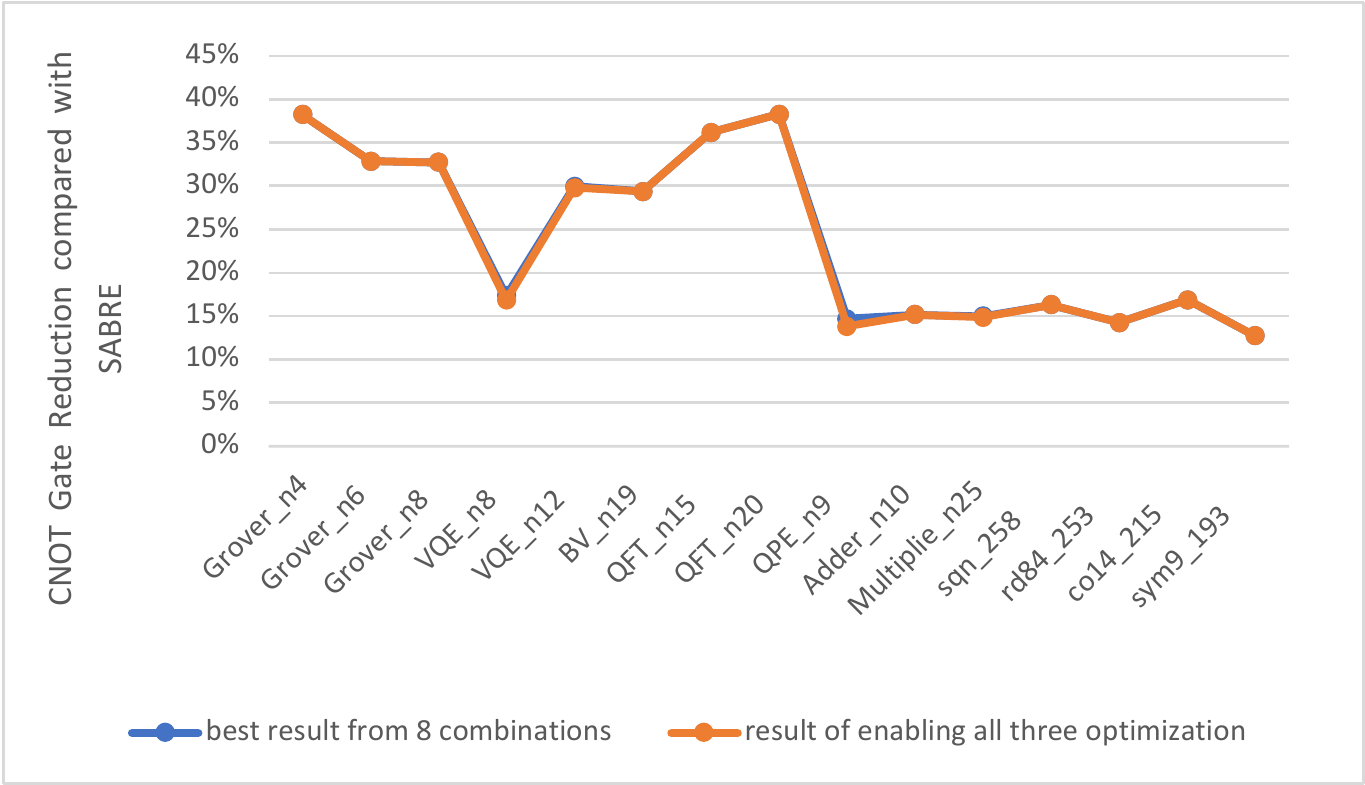}
    }\hfill
    \subfloat[\texttt{$5\times 5$ 2D grid topology}\label{fig:CNOT_reduction_compare_grid}]{  \includegraphics[width = 0.3\linewidth]{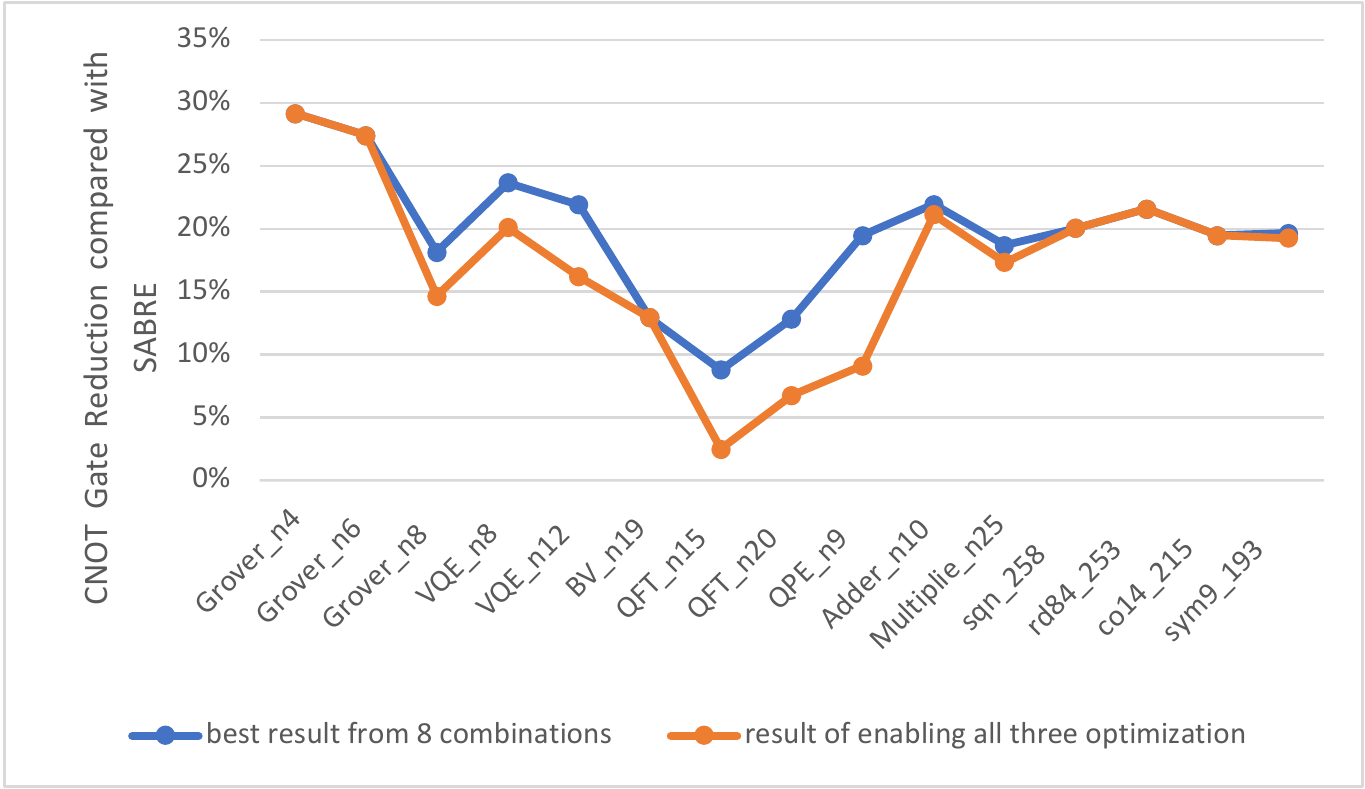}
    }
    \caption{CNOT gate reduction comparison between the best result from the 8 combinations and the result of enabling all the three optimizations on three different connectivity maps}
    \label{fig:8_comparison}
\end{figure*}

The value $C_{commute1}$ represents the CNOT gate reduction in the first scenario. $C_{commute1}$ is set to 2 when the CNOT gate cancellation happens, and set to 0 when the SWAP can not be optimized. %In the layered search scheme, there may be multiple locations to insert SWAP gates . 
In our implementation, we move the  \texttt{CommutationAnalysis} pass before routing to analyze the gate commutation relationship. The \texttt{CommutationAnalysis} pass traverses the DAG and groups the commutable gates that operate on the same qubits into the same commute set. As pointed out earlier, single-qubit gates before a SWAP should not block the commutation analysis. Therefore, our compiler skips the single-qubit gates before SWAPs to find cancellable CNOTs in the predecessors. The compiler finds the intersection of the commute sets for each qubit. If there is a CNOT gate in the intersection and it operates on the same qubits as the SWAP, this CNOT can be canceled with the CNOTs in the SWAP gate. If the compiler finds a cancellable CNOT in the predecessors, the single-qubit gates before the SWAP gate will be moved to after the SWAP gate. And the compiler will label the SWAP based on the control and the target qubit of the CNOT. After routing, the SWAP will be decomposed based on this label. In our experiments, we found the commute set size is always smaller than 20 gates. In some of the benchmarks (for example, VQE with specific ansatz designs), it is possible that all the gates commute and form very large commute sets. We limit the search size in the commute set to avoid the potentially long search time. The compiler will only search for the first 20 gates in the commute sets. Thereby, the time complexity of calculating $C_{commute1}$ and $C_{commute2}$ is $O(1)$. %If the compiler does not find a cancellable CNOT, it would perform a similar search in the predecessors of the SWAP gate. 
The compiler also merges the single-qubit gates before and after the SWAP gate. Although the $t\ket{ket}$ compiler~\cite{sivarajah2020tket} has a similar optimization \texttt{CommuteSQThroughSWAP} that commutes the single-qubit gates through SWAP, it always places the single-qubit gates on the physical qubit with the best fidelity. In our approach, the single-qubit gate placement is optimization-aware, which leads to more gate cancellation opportunities.

The value $C_{commute2}$ represents the CNOT gate reduction of the second scenario, i.e., a set of commutable gates sandwiched by two SWAP gates. $C_{commute2}$ is set to 2 when either of the SWAPs has one CNOT gate canceled. And it is set to 0 when the SWAPs are not optimized. The compiler will search for the commute set on both of the qubits of a SWAP. Similar to $C_{commute1}$, the compiler skips the single-qubit gates before a SWAP. Once the compiler finds that a commute set is sandwiched by two SWAPs, it will check if the CNOT in the SWAP might commute with the gates in the commute set. The SWAP gates are also labeled for optimization-aware decomposition.

\subsection{Optimization Selection}
\label{subsec:strategy}

In previous sections, three different optimizations are discussed. The first one is the two-qubit block re-synthesis, and the other two are the commutation-based optimizations. Since different quantum algorithms have different circuit structures, they may favor different optimizations. In other words, enabling different sets of optimizations may have different impacts on different quantum algorithms. Since we have three optimizations, enabling/disabling these three different optimizations and differently combining them will generate a total of 8 possible combinations. Figure~\ref{fig:CNOT_reduction_compare_montreal}, Figure~\ref{fig:CNOT_reduction_compare_linear} and Figure~\ref{fig:CNOT_reduction_compare_grid} show the CNOT reduction comparison between the best result of the 8 combinations and the result of enabling all the three optimizations on three different coupling maps, which are the ibmq\_montreal coupling map, the 25-qubit linear nearest neighbour coupling map, and the $5\times 5$ 2D grid coupling map, respectively. From the results, we can see that for most of the benchmarks, enabling all the three optimizations has a very close CNOT reduction ratio compared with the best result of the 8 combinations. Therefore, enabling all three optimizations is used in our NASSC approach.

\subsection{Noise-aware Qubit Routing}
\label{subsec:noise}
In this section, we discuss the noise-awareness of the NASSC algorithm. Since the gate cancellation leads to higher noise reduction, our NASSC algorithm prioritizes the gate optimizations over the noise awareness. Both SABRE and NASSC routing algorithms can be noise-aware by incorporating the noise information in the distance matrix~\cite{niu2020hardware}. The modified distance matrix considers the CNOT gate error $\varepsilon[g.i][g.j]$, SWAP execution time $T[g.i][g.j]$, and the original distance $D[g.i][g.j]$ between qubit $g.i$ and $g.j$:

\begin{equation} 
    \resizebox{0.9\linewidth}{!}{%
$\begin{aligned}
    D_{noise}[g.i][g.j] = \alpha_1\times \varepsilon[g.i][g.j] + \alpha_2\times T[g.i][g.j]\\
    + \alpha_3\times D[g.i][g.j]
\end{aligned}$
}
\end{equation}
Here $\alpha_1$, $\alpha_2$, $\alpha_3$ are the normalized parameters. $\alpha_1$, $\alpha_2$, $\alpha_3$ are set to be 0.5, 0, and 0.5 in our experiments. In Section~\ref{subsec:noisemodel}, we compare the original NASSC algorithm and the noise-aware version of NASSC algorithm on a realistic noise model and the results show that the original NASSC algorithm actually achieves the best route.

\subsection{Complexity Analysis}
\label{subsec:complexity}

In this section, we show that our NASSC algorithm has the same level of time complexity as SABRE. SABRE is known for its low time complexity $O(N^{2.5})$ compared to the exhaustive search approaches~\cite{zulehner2018efficient, wille2016look} with $O(exp(N))$ complexity for each two-qubit gate. Here N is the number of physical qubits. In the NASSC algorithm, the time complexity to satisfy each two-qubit gate is the multiplication of the time to calculate the cost function ($O(N)$), the maximum number of the SWAP candidates ($O(N)$) per iteration, and the maximum number of iterations per two-qubit gate ($O(\sqrt{N})$). As discussed in Section~\ref{subsec:costfunction}, evaluating the cost function for a SWAP candidate has complexity $O(N)$. In each iteration, the maximum number of SWAP candidates is linearly correlated with the size of the front layer which is $O(N)$. The maximum number of iterations per two-qubit gate is the diameter of the device coupling graph ($O(\sqrt{N})$ for 2D graph). In other words, a two-qubit gate needs at most $O(\sqrt{N})$ SWAP gates to move two qubits together. So the total time complexity to satisfy each two-qubit gate is $O(N^{2.5})$. Note that the two-qubit re-synthesis pass and the commutation analysis pass are predefined passes that would be executed at the gate optimization step. Therefore, moving these two passes before routing does not incur extra complexity.

\begin{figure}
    \centering
    \subfloat[The connectivity map of 27-qubit \texttt{ibmq\_montreal} \label{subfig:montreal}]{  \includegraphics[width = 0.8\linewidth]{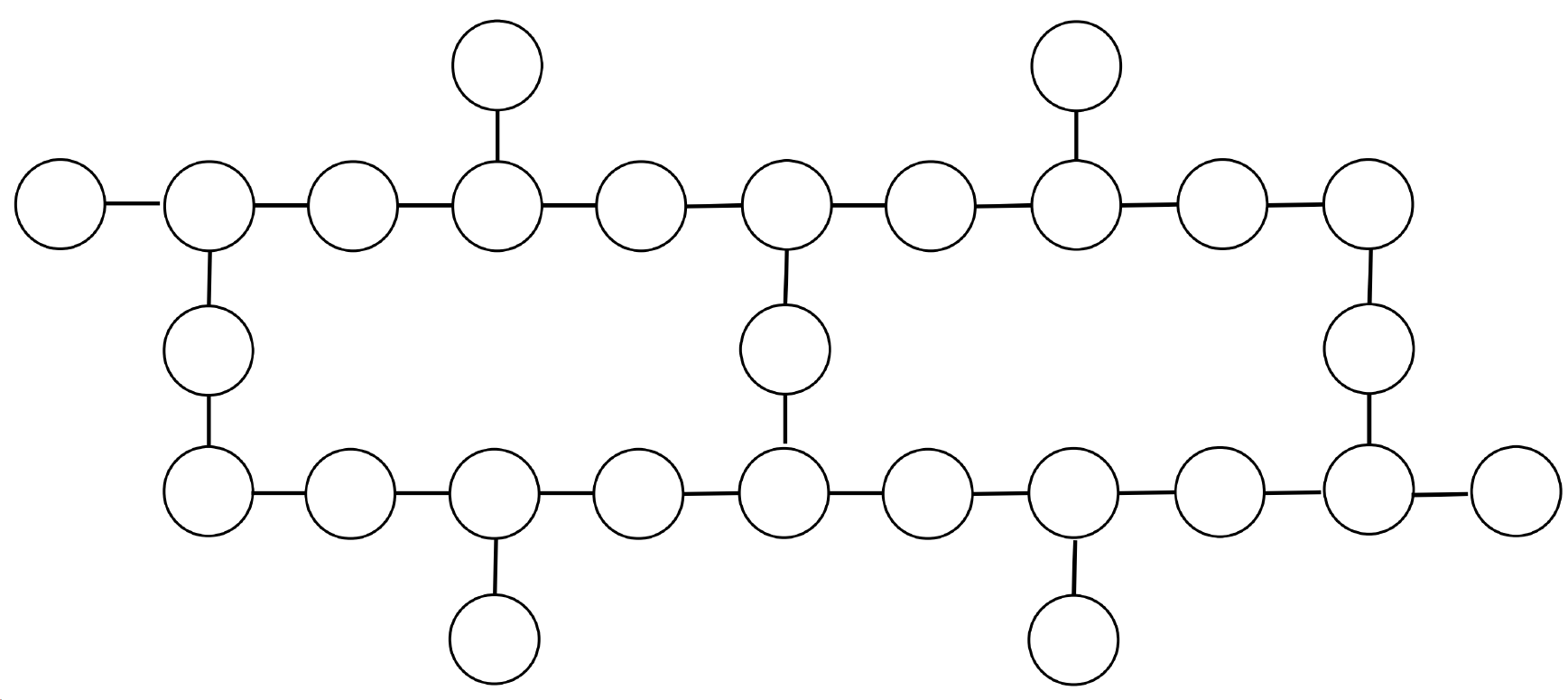}
    }\hfill
    \subfloat[The connectivity map of 25-qubit linear nearest neighbour topology\label{subfig:lnn}]{  \includegraphics[width = 0.4\linewidth]{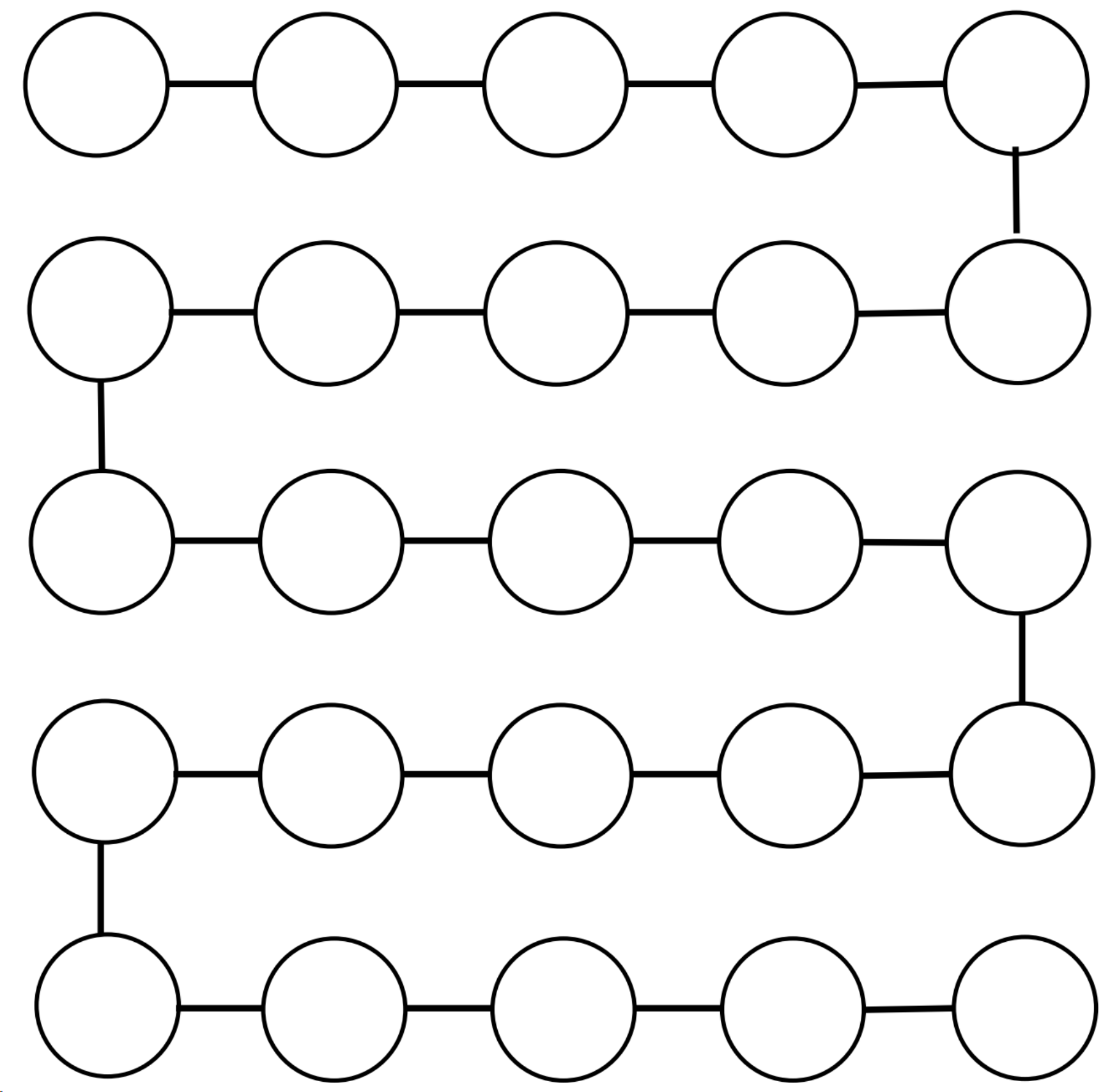}
    }\hfill
    \subfloat[The connectivity map of $5\times 5$ 2D grid topology\label{subfig:2dgrid}]{  \includegraphics[width = 0.4\linewidth]{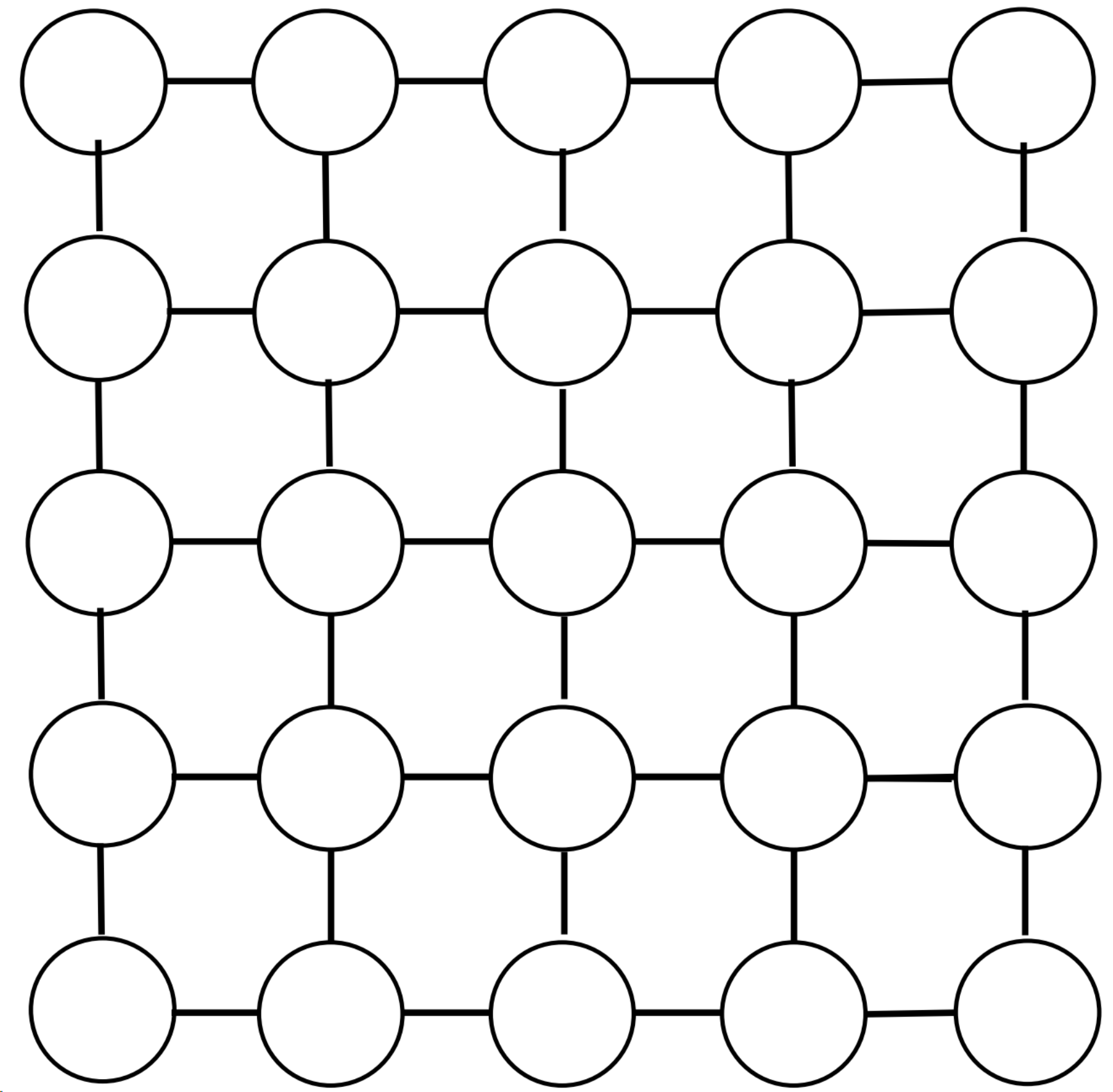}
    }
    \caption{Three different connectivity maps}
    \label{fig:connectivity_map}
\end{figure}
\section{Methodology}
\label{sec:method}

We evaluate NASSC with a set of benchmarks on devices with different connectivity topologies.

\textbf{Benchmarks:} The benchmarks are derived from the algorithms in ~\cite{nielsen2002quantum}, Qiskit programs~\cite{Qiskit}, QASMBench~\cite{li2020qasmbench} and RevLib~\cite{4539430}.

\textbf{Algorithm Implementation:} We implement our NASSC algorithm on the open-source quantum computing framework Qiskit~\cite{Qiskit}, the version of qiskit-terra is 0.19.0, and our implementation is publicly available (see Appendix). %~\footnote{hidden for double blinded review}. SABRE is one of Qiskit's preset routing algorithm. 
In our experiments we compared the Qiskit with the SABRE routing algorithm (Qiskit+SABRE) and the Qiskit with our NASSC algorithm (Qiskit+NASSC).

\textbf{Algorithm Configuration and Evaluation:} In our experiment, the weight of the extended layer $E$ is set to 0.5, and the size of the layer $|E|$ is 20. The layer size represents the maximum number of two-qubit gates in the layer. The experiments with the SABRE algorithm use the same extended layer size and weight. All the binary values $b_k$ are set to 1 to enable all the optimizations. We use the geometric mean to calculate the average ratio of the CNOT gate and depth reduction. The results are the average of ten runs.

\textbf{Connectivity:} We run our experiments with three different hardware connectivity maps, including the connectivity of a 27-qubit quantum computer \texttt{ibmq\_montreal}, linear nearest neighbour, and 2D grid. Figure~\ref{fig:connectivity_map} shows the different connectivity maps in our experiments. The \texttt{ibmq\_montreal} has heavy-hex lattice topology that might be used by IBM for future large scale quantum computers with error correction code~\cite{IBMheavyhex}.

\section{Evaluation}
\label{sec:evaluation}
In this section, we first compare the CNOT gate number and circuit depths of using our NASSC approach with SABRE on the coupling map of IBM quantum device \texttt{ibmq\_montreal}. Then, we analyze the transpilation time for large-size quantum circuits. Next, we compare the CNOT gate number of using our NASSC approach with SABRE on different connectivity maps. Finally, we run experiments using the Qiskit simulator with noise model from real quantum device \texttt{ibmq\_montreal} to show the success rate improvement when applying our NASSC approach.

\subsection{Comparison with Qiskit+SABRE}
We compare the number of additional CNOTs and the additional circuit depth of our NASSC with Qiskit+SABRE on the coupling map of IBM quantum device \texttt{ibmq\_montreal}. The CNOT gate reductions are shown in Table~\ref{table:result_montreal}. And the depth reductions are shown in Table~\ref{table:depth_montreal_result}. The $\text{CNOT}_{total}$ in the table means the total number of CNOTs in the original circuit optimized by Qiskit. The $\text{CNOT}_{total}$ under Qiskit+SABRE/Qiskit+NASSC column means the total number of CNOTs after SABRE/NASSC routing and Qiskit optimizations. The $\text{CNOT}_{add}$ under Qiskit+SABRE/Qiskit+NASSC column means the number of additional CNOTs after SABRE/NASSC routing and Qiskit optimizations compared with the original circuit optimized by Qiskit. $\Delta \text{CNOT}_{total}$ is the percentage change in the number of total CNOT gates: $\Delta \text{CNOT}_{total}= 1 - \text{CNOT}_{total}(\text{NASSC})/\text{CNOT}_{total}(\text{SABRE})$. $\Delta \text{CNOT}_{add}$ is the percentage change in the number of additional CNOT gates: $\Delta \text{CNOT}_{add}= 1 - \text{CNOT}_{add}(\text{NASSC})/\text{CNOT}_{add}(\text{SABRE})$. For all of the benchmarks, the additional CNOTs of Qiskit+NASSC are less than that of Qiskit+SABRE. The geometric means of $\Delta \text{CNOT}_{total}$ and $\Delta \text{CNOT}_{add}$ are $13.25\%$ and $21.30\%$. Since the NASSC algorithm also incorporates re-synthesis, we compare the total transpilation time of Qiskit+NASSC with Qiskit+SABRE, which is collected by transpiling each benchmark ten times and getting the average value. As shown in Table~\ref{table:result_montreal}, the transpilation time of our NASSC is comparable to that of SABRE. We also perform the circuit depth comparison in Table~\ref{table:depth_montreal_result}. For most of the benchmarks, the circuit depth of Qiskit+NASSC is reduced compared with Qiskit+SABRE. The reason is that at the routing step, our algorithm will try to merge the single-qubit gates before and after the SWAP gate. This single-qubit movement also contributes to the circuit depth reduction. However, for a few (5 out of 15) of the benchmarks, the circuit depth is increased. This is due to the fact that the re-synthesis of the two-qubit blocks may generate more single-qubit gates, which would increase the depth of the circuits. Note that in state-of-the-art quantum systems, the two-qubit gates' error rates are much higher than that of the single-qubit gates. As a result, it is often desirable to have reduced numbers of 2-qubit gates even at the cost of higher circuit depths due to additional single-qubit gates.

\begin{table*}[bhtp]
  \caption{Number of additional CNOT gates of NASSC in comparison with Qiskit+SABRE~\cite{li2019sabre} on \texttt{ibmq\_montreal} (The geometric mean of $\Delta CNOT_{total}$ and $\Delta CNOT_{add}$ is 13.25\% and 21.30\%, respectively)}
  \label{table:result_montreal}
  \centering
  \small
  \resizebox{\linewidth}{!}{%
    \begin{threeparttable}
  \begin{tabular}{|c|c|c|c|c|c|c|c|c|c|c|c|}
  \hline
  \multicolumn{3}{|c|}{Original Circuit} &  \multicolumn{3}{c|}{Qiskit+SABRE} &  \multicolumn{3}{c|}{Qiskit+NASSC} &  \multicolumn{3}{c|}{Comparison}\\
    \hline
     name & $\#qubits$ & $CNOT_{total}$ & $CNOT_{total}$ & $CNOT_{add}$ & transpile time(s) & $CNOT_{total}$ & $CNOT_{add}$ & transpile time(s) & $\Delta CNOT_{total}$ & $\Delta CNOT_{add}$ & $t_\text{NASSC}/t_\text{SABRE}$\\
    \hline
Grover 4-qubits      & 4        & 84                           & 167                       & 83                      & 1.35                          & 116                       & 32                      & 1.75                          & 30.54\%            & 61.45\%          & 1.30              \\\hline
Grover 6-qubits      & 6        & 184                          & 310                       & 126                     & 2.43                          & 231                       & 47                      & 3.2                           & 25.48\%            & 62.70\%          & 1.32              \\\hline
Grover 8-qubits      & 8        & 760                          & 1470                      & 710                     & 10.15                         & 978                       & 218                     & 12.4                          & 33.47\%            & 69.30\%          & 1.22              \\\hline
VQE 8-qubits         & 8        & 84                           & 370                       & 286                     & 2.92                          & 290                       & 206                     & 4.29                          & 21.62\%            & 27.97\%          & 1.47              \\\hline
VQE 12-qubits        & 12       & 198                          & 919                       & 721                     & 7.42                          & 790                       & 592                     & 9.28                          & 14.04\%            & 17.89\%          & 1.25              \\\hline
BV 19-qubits         & 19       & 18                           & 134                       & 116                     & 0.75                          & 132                       & 114                     & 1.03                          & 1.49\%             & 1.72\%           & 1.38              \\\hline
QFT 15-qubits        & 15       & 210                          & 591                       & 381                     & 3.51                          & 434                       & 224                     & 6.01                          & 26.57\%            & 41.21\%          & 1.71              \\\hline
QFT 20-qubits        & 20       & 374                          & 1014                      & 640                     & 6.44                          & 901                       & 527                     & 10.81                         & 11.14\%            & 17.66\%          & 1.68              \\\hline
QPE 9-qubits         & 9        & 43                           & 114                       & 71                      & 0.81                          & 102                       & 59                      & 1.39                          & 10.53\%            & 16.90\%          & 1.72              \\\hline
Adder 10-qubits      & 10       & 65                           & 186                       & 121                     & 1.18                          & 160                       & 95                      & 1.74                          & 13.98\%            & 21.49\%          & 1.48              \\\hline
Multiplier 25-qubits & 25       & 670                          & 1889                      & 1219                    & 15.7                          & 1836                      & 1166                    & 20.47                         & 2.81\%             & 4.35\%           & 1.30              \\\hline
sqn\_258             & 10       & 4459                         & 11939                     & 7480                    & 109.35                        & 10194                     & 5735                    & 118.35                        & 14.62\%            & 23.33\%          & 1.08              \\\hline
rd84\_253            & 12       & 5960                         & 16006                     & 10046                   & 149.97                        & 13399                     & 7439                    & 159.13                        & 16.29\%            & 25.95\%          & 1.06              \\\hline
co14\_215            & 15       & 7840                         & 19788                     & 11948                   & 190.9                         & 17242                     & 9402                    & 210.62                        & 12.87\%            & 21.31\%          & 1.10              \\\hline
sym9\_193            & 11       & 15232                        & 40189                     & 24957                   & 379.8                         & 34014                     & 18782                   & 389.1                         & 15.36\%            & 24.74\%          & 1.02              \\\hline
\multicolumn{1}{|c|}{Geometric mean} & \multicolumn{8}{c|}{}                                                                            & \multicolumn{1}{c|}{13.25\%} & \multicolumn{1}{c|}{21.30\%}& \multicolumn{1}{c|}{1.32} \\ \hline

  \end{tabular}
      \begin{tablenotes}
      \large
      \item $\Delta \text{CNOT}_{total}$ is the percentage change in the number of total CNOT gates: $\Delta \text{CNOT}_{total}= 1 - \text{CNOT}_{total}(\text{NASSC})/\text{CNOT}_{total}(\text{SABRE})$.
      \item $\Delta \text{CNOT}_{add}$ is the percentage change in the number of additional CNOT gates: $\Delta \text{CNOT}_{add}= 1 - \text{CNOT}_{add}(\text{NASSC})/\text{CNOT}_{add}(\text{SABRE})$.
      \item $t_\text{NASSC}/t_\text{SABRE}$ is the ratio between the total transpilation time of Qiskit+NASSC and Qiskit+SABRE: $t_\text{NASSC}/t_\text{SABRE} = \text{transpile time(NASSC)}/\text{transpile time(SABRE)}$
    \end{tablenotes}
  \end{threeparttable}
  }
\end{table*}

\begin{table*}[bhtp]
  \caption{Circuit depth of NASSC in comparison with Qiskit+SABRE~\cite{li2019sabre} on \texttt{ibmq\_montreal} (The geometric mean of $\Delta depth_{total}$ and $\Delta depth_{add}$ is 6.05\% and 7.61\%, respectively)}
  \label{table:depth_montreal_result}
  \centering
  \small
    \resizebox{\linewidth}{!}{%
      \begin{threeparttable}
  \begin{tabular}{|c|c|c|c|c|c|c|c|c|}
  \hline
  \multicolumn{3}{|c|}{Original Circuit} &  \multicolumn{2}{c|}{Qiskit+SABRE} &  \multicolumn{2}{c|}{Qiskit+NASSC} &  \multicolumn{2}{c|}{Comparison}\\
    \hline
     name & $\#qubits$ & $depth_{total}$ & $depth_{total}$ & $depth_{add}$ & $depth_{total}$ & $depth_{add}$ & $\Delta depth_{total}$ & $\Delta depth_{add}$\\
    \hline
Grover 4-qubits      & 4        & 155                           & 286                        & 131                      & 284                        & 129                      & $0.70\%$              & $1.53\%$            \\\hline
Grover 6-qubits      & 6        & 315                           & 512                        & 197                      & 444                        & 129                      & $13.28\%$             & $34.52\%$           \\\hline
Grover 8-qubits      & 8        & 1275                          & 2010                       & 735                      & 1724                       & 449                      & $14.23\%$             & $38.91\%$           \\\hline
VQE 8-qubits         & 8        & 108                           & 356                        & 248                      & 345                        & 237                      & $3.09\%$              & $4.44\%$            \\\hline
VQE 12-qubits        & 12       & 171                           & 719                        & 548                      & 607                        & 436                      & $15.58\%$             & $20.44\%$           \\\hline
BV 19-qubits         & 19       & 25                            & 85                         & 60                       & 103                        & 78                       & $-21.18\%$            & $-30.00\%$          \\\hline
QFT 15-qubits        & 15       & 100                           & 407                        & 307                      & 321                        & 221                      & $21.13\%$             & $28.01\%$           \\\hline
QFT 20-qubits        & 20       & 135                           & 584                        & 449                      & 594                        & 459                      & $-1.71\%$             & $-2.23\%$           \\\hline
QPE 9-qubits         & 9        & 84                            & 148                        & 64                       & 170                        & 86                       & $-14.86\%$            & $-34.38\%$          \\\hline
Adder 10-qubits      & 10       & 119                           & 232                        & 113                      & 242                        & 123                      & $-4.31\%$             & $-8.85\%$           \\\hline
Multiplier 25-qubits & 25       & 799                           & 1832                       & 1033                     & 2157                       & 1358                     & $-17.74\%$            & $-31.46\%$          \\\hline
sqn\_258             & 10       & 6013                          & 19356                      & 13343                    & 14642                      & 8629                     & $24.35\%$             & $35.33\%$           \\\hline
rd84\_253            & 12       & 8011                          & 25971                      & 17960                    & 19330                      & 11319                    & $25.57\%$             & $36.98\%$           \\\hline
co14\_215            & 15       & 9490                          & 30330                      & 20840                    & 23699                      & 14209                    & $21.86\%$             & $31.82\%$           \\\hline
sym9\_193            & 11       & 21332                         & 68008                      & 46676                    & 47704                      & 26372                    & $29.86\%$             & $43.50\%$           \\\hline
\multicolumn{1}{|c|}{Geometric mean} & \multicolumn{6}{c|}{}                                                                            & \multicolumn{1}{c|}{$6.05\%$} & \multicolumn{1}{c|}{$7.61\%$} \\ \hline

  \end{tabular}
      \begin{tablenotes}
      \footnotesize
      \item $\Delta\text{depth}_{total}$ is the percentage change in total circuit depth: $\Delta \text{depth}_{total}$=$1-\text{depth}_{total}(\text{NASSC})/\text{depth}_{total}(\text{SABRE})$.
      \item $\Delta\text{depth}_{add}$ is the percentage change in additional circuit depth: $\Delta\text{depth}_{add}$=$1-\text{depth}_{add}(\text{NASSC})/\text{depth}_{add}(\text{SABRE})$.
    \end{tablenotes}
  \end{threeparttable}
  }
\end{table*}

\begin{table*}[bhtp]
  \caption{Number of additional CNOT gates of NASSC in comparison with Qiskit+SABRE~\cite{li2019sabre} on \texttt{linear topology} (The geometric mean of $\Delta CNOT_{total}$ and $\Delta CNOT_{add}$ is 21.92\% and 34.65\%, respectively)}
  \label{table:lnn_results}
  \centering
  \small
    \resizebox{\linewidth}{!}{%
  \begin{tabular}{|c|c|c|c|c|c|c|c|c|c|c|c|}
  \hline
  \multicolumn{3}{|c|}{Original Circuit} &  \multicolumn{3}{c|}{Qiskit+SABRE} &  \multicolumn{3}{c|}{Qiskit+NASSC} &  \multicolumn{3}{c|}{Comparison}\\
    \hline
     name & $\#qubits$ & $CNOT_{total}$ & $CNOT_{total}$ & $CNOT_{add}$ & transpile time(s) & $CNOT_{total}$ & $CNOT_{add}$ & transpile time(s) & $\Delta CNOT_{total}$ & $\Delta CNOT_{add}$ & $t_\text{NASSC}/t_\text{SABRE}$\\
    \hline
Grover 4-qubits      & 4        & 84                           & 238                       & 154                     & 1.69                          & 147                       & 63                      & 2.56                          & 38.24\%            & 59.09\%          & 1.52              \\\hline
Grover 6-qubits      & 6        & 184                          & 414                       & 230                     & 2.5                           & 278                       & 94                      & 3.94                          & 32.85\%            & 59.13\%          & 1.58              \\\hline
Grover 8-qubits      & 8        & 760                          & 1867                      & 1107                    & 10.26                         & 1256                      & 496                     & 15.46                         & 32.73\%            & 55.19\%          & 1.51              \\\hline
VQE 8-qubits         & 8        & 84                           & 202                       & 118                     & 2.75                          & 168                       & 84                      & 5.59                          & 16.83\%            & 28.81\%          & 2.03              \\\hline
VQE 12-qubits        & 12       & 198                          & 564                       & 366                     & 6.13                          & 396                       & 198                     & 13.45                         & 29.79\%            & 45.90\%          & 2.19              \\\hline
BV 19-qubits         & 19       & 18                           & 269                       & 251                     & 1.33                          & 190                       & 172                     & 1.64                          & 29.37\%            & 31.47\%          & 1.24              \\\hline
QFT 15-qubits        & 15       & 210                          & 492                       & 282                     & 3.11                          & 314                       & 104                     & 9.82                          & 36.18\%            & 63.12\%          & 3.16              \\\hline
QFT 20-qubits        & 20       & 374                          & 922                       & 548                     & 5.23                          & 569                       & 195                     & 18.59                         & 38.29\%            & 64.42\%          & 3.55              \\\hline
QPE 9-qubits         & 9        & 43                           & 116                       & 73                      & 0.84                          & 100                       & 57                      & 1.38                          & 13.79\%            & 21.92\%          & 1.63              \\\hline
Adder 10-qubits      & 10       & 65                           & 132                       & 67                      & 0.91                          & 112                       & 47                      & 1.39                          & 15.15\%            & 29.85\%          & 1.53              \\\hline
Multiplier 25-qubits & 25       & 670                          & 2190                      & 1520                    & 13.07                         & 1865                      & 1195                    & 20.82                         & 14.84\%            & 21.38\%          & 1.59              \\\hline
sqn\_258             & 10       & 4459                         & 12286                     & 7827                    & 112.32                        & 10282                     & 5823                    & 115.32                        & 16.31\%            & 25.60\%          & 1.03              \\\hline
rd84\_253            & 12       & 5960                         & 17926                     & 11966                   & 157.39                        & 15373                     & 9413                    & 163.89                        & 14.24\%            & 21.34\%          & 1.04              \\\hline
co14\_215            & 15       & 7840                         & 22734                     & 14894                   & 207.14                        & 18897                     & 11057                   & 212.1                         & 16.88\%            & 25.76\%          & 1.02              \\\hline
sym9\_193            & 11       & 15232                        & 44418                     & 29186                   & 395.2                         & 38770                     & 23538                   & 411.14                        & 12.72\%            & 19.35\%          & 1.04              \\\hline
\multicolumn{1}{|c|}{Geometric mean} & \multicolumn{8}{c|}{}                                                                            & \multicolumn{1}{c|}{21.92\%} & \multicolumn{1}{c|}{34.65\%} & \multicolumn{1}{c|}{1.59}\\ \hline

  \end{tabular}
  }
\end{table*}

\begin{table*}[bhtp]
  \caption{Number of additional CNOT gates of NASSC in comparison with Qiskit+SABRE~\cite{li2019sabre} on \texttt{2D grid topology} (The geometric mean of $\Delta CNOT_{total}$ and $\Delta CNOT_{add}$ is 15.13\% and 28.10\%, respectively)}
  \label{table:2d_results}
  \centering
  \small
    \resizebox{\linewidth}{!}{%
  \begin{tabular}{|c|c|c|c|c|c|c|c|c|c|c|c|}
  \hline
  \multicolumn{3}{|c|}{Original Circuit} &  \multicolumn{3}{c|}{Qiskit+SABRE} &  \multicolumn{3}{c|}{Qiskit+NASSC} &  \multicolumn{3}{c|}{Comparison}\\
    \hline
     name & $\#qubits$ & $CNOT_{total}$ & $CNOT_{total}$ & $CNOT_{add}$ & transpile time(s) & $CNOT_{total}$ & $CNOT_{add}$ & transpile time(s) & $\Delta CNOT_{total}$ & $\Delta CNOT_{add}$ & $t_\text{NASSC}/t_\text{SABRE}$\\
    \hline
Grover 4-qubits      & 4        & 84                           & 161                       & 77                      & 1.4                           & 114                       & 30                      & 1.69                          & 29.19\%            & 61.04\%          & 1.20              \\\hline
Grover 6-qubits      & 6        & 184                          & 321                       & 137                     & 1.95                          & 233                       & 49                      & 3.64                          & 27.41\%            & 64.23\%          & 1.87              \\\hline
Grover 8-qubits      & 8        & 760                          & 1119                      & 359                     & 6.61                          & 955                       & 195                     & 14.72                         & 14.66\%            & 45.68\%          & 2.23              \\\hline
VQE 8-qubits         & 8        & 84                           & 224                       & 140                     & 1.81                          & 179                       & 95                      & 3.48                          & 20.09\%            & 32.14\%          & 1.93              \\\hline
VQE 12-qubits        & 12       & 198                          & 525                       & 327                     & 4.58                          & 440                       & 242                     & 7.57                          & 16.19\%            & 25.99\%          & 1.65              \\\hline
BV 19-qubits         & 19       & 18                           & 116                       & 98                      & 0.7                           & 101                       & 83                      & 1.07                          & 12.93\%            & 15.31\%          & 1.52              \\\hline
QFT 15-qubits        & 15       & 210                          & 445                       & 235                     & 3.06                          & 434                       & 224                     & 6.48                          & 2.47\%             & 4.68\%           & 2.12              \\\hline
QFT 20-qubits        & 20       & 374                          & 802                       & 428                     & 5.51                          & 748                       & 374                     & 10.93                         & 6.73\%             & 12.62\%          & 1.98              \\\hline
QPE 9-qubits         & 9        & 43                           & 77                        & 34                      & 0.64                          & 70                        & 27                      & 1.23                          & 9.09\%             & 20.59\%          & 1.94              \\\hline
Adder 10-qubits      & 10       & 65                           & 123                       & 58                      & 0.99                          & 97                        & 32                      & 1.47                          & 21.14\%            & 44.83\%          & 1.49              \\\hline
Multiplier 25-qubits & 25       & 670                          & 1709                      & 1039                    & 11.13                         & 1413                      & 743                     & 22.68                         & 17.32\%            & 28.49\%          & 2.04              \\\hline
sqn\_258             & 10       & 4459                         & 10551                     & 6092                    & 89.28                         & 8292                      & 3833                    & 108.55                        & 21.41\%            & 37.08\%          & 1.22              \\\hline
rd84\_253            & 12       & 5960                         & 14559                     & 8599                    & 125.38                        & 11449                     & 5489                    & 149.14                        & 21.36\%            & 36.17\%          & 1.19              \\\hline
co14\_215            & 15       & 7840                         & 19451                     & 11611                   & 166.81                        & 15855                     & 8015                    & 240.38                        & 18.49\%            & 30.97\%          & 1.44              \\\hline
sym9\_193            & 11       & 15232                        & 37239                     & 22007                   & 320.96                        & 29216                     & 13984                   & 441.71                        & 21.54\%            & 36.46\%          & 1.38              \\\hline
\multicolumn{1}{|c|}{Geometric mean} & \multicolumn{8}{c|}{}                                                                            & \multicolumn{1}{c|}{15.13\%} & \multicolumn{1}{c|}{28.10\%} & \multicolumn{1}{c|}{1.64 }\\ \hline

  \end{tabular}
  }
\end{table*}
\subsection{Transpilation Time for Large-Size Circuits}

The optimization awareness of our NASSC algorithm reduces the total circuit size and benefits the subsequent optimization passes. Our algorithm is especially suitable for large-size circuits because there can be more optimization opportunities when the circuit becomes larger. Table~\ref{table:result_montreal} and Table~\ref{table:depth_montreal_result} show that the large-size benchmarks sqn\_258, rd84\_253, co14\_215 and sym9\_193 have high CNOT gate number reduction and circuits depths reduction from NASSC compared with Qiskit+SABRE. The $\Delta \text{CNOT}_{add}$ of sqn\_258, rd84\_253, co14\_215 and sym9\_193 are 23.33\%, 25.95\%, 21.31\% and 24.74\%, respectively, and $\Delta \text{depth}_{add}$ of sqn\_258, rd84\_253, co14\_215 and sym9\_193 are 35.33\%, 36.98\%, 31.82\% and 43.50\%, respectively. The high reduction in the circuit size and circuit depth leads to shorter circuit optimization and gate scheduling time. Therefore, the total transpilation time of these large-size benchmarks only show a small increase than the SABRE algorithm. The transpilation time of sqn\_258, rd84\_253, co14\_215 and sym9\_193 compared with Qiskit+SABRE approach are only 1.08X, 1.06X, 1.10X and 1.02X, respectively.

\subsection{Backend Connectivity}

Table~\ref{table:result_montreal}, ~\ref{table:lnn_results}, and ~\ref{table:2d_results} show the number of CNOT gates using our NASSC approach in comparison with Qiskit+SABRE on different coupling maps. As shown in the tables, our NASSC approach is effective on different topologies and has high CNOT gate number reduction compared with Qiskit+SABRE on all three different coupling maps. The geometric means of $\Delta \text{CNOT}_{total}$ on \texttt{ibmq\_montreal}, linear topology, and 2D grid topology are 13.25\%, 21.92\% and 15.13\%. And the geometric means of $\Delta \text{CNOT}_{add}$ on \texttt{ibmq\_montreal}, linear topology, and 2D grid topology are 21.30\%, 34.65\% and 28.10\%. Among the three coupling maps, our NASSC approach has the highest CNOT gate number reduction compared with Qiskit+SABRE on the linear coupling map. The reason is that linear coupling map has the worst connectivity among all these three coupling maps. Therefore, there are more optimization opportunities than the other two types of coupling maps.

\subsection{Simulations using A Realistic Noise Model} 
\label{subsec:noisemodel}
\begin{figure}
    \centering
    \subfloat[Comparison of additional CNOT gate count \label{fig:addtion_CNOT_compare_montreal}]{  \includegraphics[width = 0.45\linewidth]{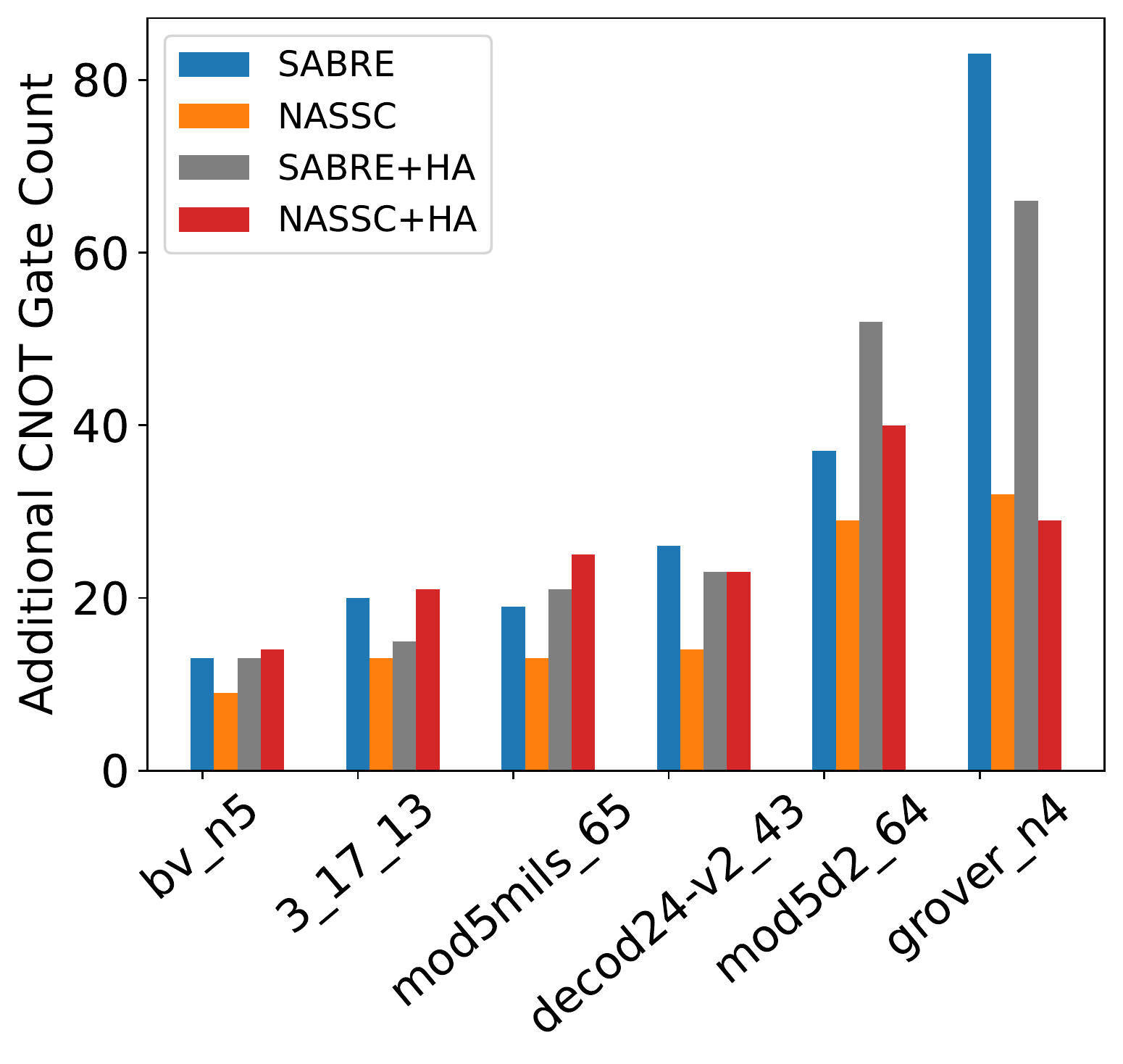}
    }\hfill
    \subfloat[Comparison of success rate \label{fig:successrate-comparison}]{
    \includegraphics[width = 0.45\linewidth]{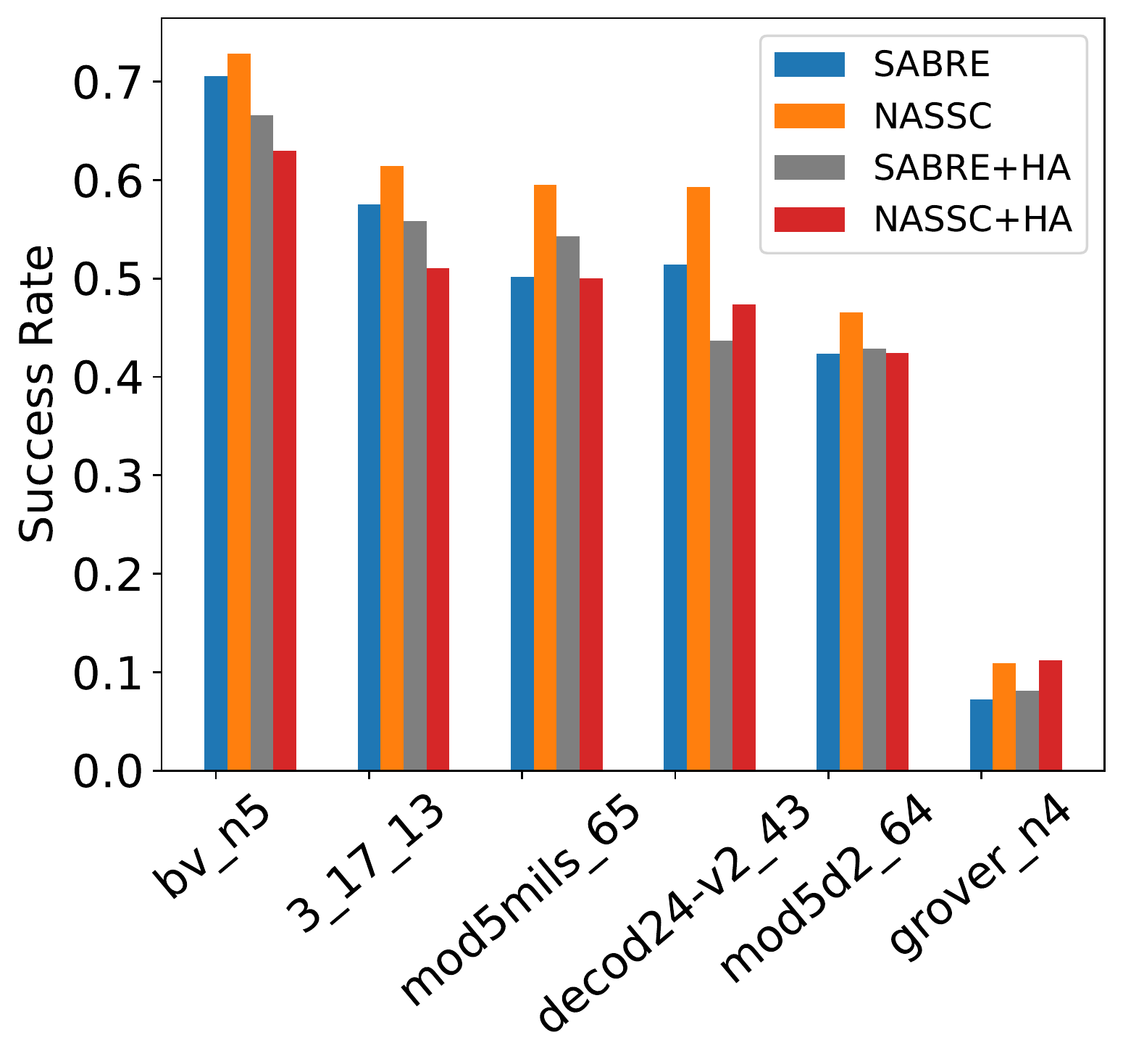}
    }
    \caption{Comparison of four different routing algorithms using noise model from the real IBM quantum device \texttt{ibmq\_montreal}}
    \label{fig:noise-comparison}
\end{figure}

We compare the results of noise-aware routing algorithms using Qiskit's simulator (8192 trials) and the noise model is generated from the properties of the real IBM quantum device \texttt{ibmq\_montreal}. The results are shown in Figure~\ref{fig:noise-comparison}. Figure~\ref{fig:addtion_CNOT_compare_montreal} shows additional CNOT gate count after applying different routing method. Figure~\ref{fig:successrate-comparison} shows the success rate after applying different routing methods. And the success rate of each benchmark means the ratio of the number of trials that get the correct output state over the total number of trials. The SABRE+HA and NASSC+HA are the noise-aware versions of the SABRE and the NASSC algorithm by modifying the distance matrix based on the calibration data from the real IBM quantum device \texttt{ibmq\_montreal}. Among the four routing algorithms, the NASSC algorithm has the fewest additional CNOT gate count and lowest noise.

\section{Conclusions}
\label{sec:conclusions}
In this paper, we propose an optimization-aware qubit routing algorithm, NASSC. We highlight that optimization-aware routing leads to better routing decisions and benefits subsequent optimizations. We evaluate different optimization combinations at the routing step. Our experiments show that the routing overhead compiled with our routing algorithm is significantly reduced in terms of the number of CNOT gates and circuit depths.

\section*{Acknowledgment}
We thank the anonymous reviewers for their valuable comments. This work is funded in part by NSF grants 1717550, 1908406, 1818914 (with subcontract to NC State University from Duke University) and 2120757 (with subcontract to NC State University from University of Maryland). %\todo



%\bibliographystyle{IEEEtran}

\newpage
\appendix
\section{Artifact Appendix}

%%%%%%%%%%%%%%%%%%%%%%%%%%%%%%%%%%%%%%%%%%%%%%%%%%%%%%%%%%%%%%%%%%%%%
\subsection{Abstract}
Our artifact provides the source code of the optimization-aware qubit routing algorithm NASSC, which is developed based on the open-source framework Qiskit-Terra. All the benchmarks that are used in the evaluation part are also included in the artifact.  Python scripts are used to run all these benchmarks and generate the corresponding results of the evaluation part.

\subsection{Artifact check-list (meta-information)}

{\small
\begin{itemize}
  \item {\bf Algorithm: } NASSC routing algorithm
  \item {\bf Compilation: } Qiskit Terra transpiler
  \item {\bf Data set: } Benchmarks listed in Section~\ref{sec:evaluation} of our paper
  \item {\bf Hardware: } In Section~\ref{subsec:noisemodel}, simulations are done based on the noise model from the real IBM quantum device \texttt{ibmq\_montreal}
  \item {\bf Execution: } Run the bash scripts and python scripts
  \item {\bf Metrics: } CNOT gate count, circuit depth, transpilation time, and success rate of the quantum circuit
  \item {\bf Output: } CSV files and pdf files corresponding to the results of tables and figures in Section~\ref{sec:evaluation} of our paper
  \item {\bf Experiments: } Applying different routing method to the quantum circuits and compare the CNOT gate number, circuit depth and transpilation time of the circuit
  \item {\bf How much disk space required (approximately)?: } 2GB
  \item {\bf How much time is needed to prepare workflow (approximately)?: } A few minutes
  \item {\bf How much time is needed to complete experiments (approximately)?: } 12 hours
  \item {\bf Publicly available?: } Yes
  \item {\bf Code licenses (if publicly available)?: } Apache-2.0 License
  \item {\bf Archived (provide DOI)?: } 10.5281/zenodo.5790219
\end{itemize}
}

%%%%%%%%%%%%%%%%%%%%%%%%%%%%%%%%%%%%%%%%%%%%%%%%%%%%%%%%%%%%%%%%%%%%%
\subsection{Description}

\subsubsection{How to access}

Our source code, benchmarks and the scripts to run the benchmarks are available on Zenodo: https://doi.org/10.5281/zenodo.5790219. We also provide a github repository for potential updated versions: https://github.com/peiyi1/nassc\_code

\subsubsection{Hardware dependencies}
In Section~\ref{subsec:noisemodel} of our paper, we access a real IBM quantum device \texttt{ibmq\_montreal} to obtain the noise model from it, and then perform the simulations based on that noise model. If you do not have access to the real \texttt{ibmq\_montreal}, we have provided the experiments using the noise model from the fake \texttt{ibmq\_montreal} which do not need access to a real quantum device.

\subsubsection{Software dependencies}
Python version used in the experiments is 3.7, and all the experiments have been tested in Red Hat Enterprise Linux Server 7.9 and Ubuntu 18.04.6 LTS.

\subsubsection{Data sets}
Quantum benchmarks are listed in our paper.

%%%%%%%%%%%%%%%%%%%%%%%%%%%%%%%%%%%%%%%%%%%%%%%%%%%%%%%%%%%%%%%%%%%%%
\subsection{Installation}

\subsubsection{Anaconda installation}
Anaconda can be downloaded in https://www.anaconda.com/. After installing Anaconda, create an environment named env:

\$ conda create -y -n env python=3.7

Then activate the environment:

\$ conda activate env \\

\subsubsection{Qiskit installation}
Download our repository from Zenodo: https://doi.org/10.5281/zenodo.5790219

After downloading the repository from Zenodo, enter the folder of nassc\_code, there are two folders named qiskit-terra and qiskit-ibmq-provider. Let's first enter the folder of qiskit-terra and install it using the following commands:

\$ pip install cython

\$ pip install -r requirements-dev.txt

\$ pip install .

After qiskit-terra is installed, go to the folder of qiskit-ibmq-provider and install it using the following commands:

\$ pip install -r requirements-dev.txt

\$ pip install . \\

\subsubsection{Package installation}
Before running the experiments, go back to the folder of nassc\_code and install the benchmark package and the hamap package using the following commands:

\$ python setup\_benchmark.py develop

\$ python setup\_hamap.py develop

%%%%%%%%%%%%%%%%%%%%%%%%%%%%%%%%%%%%%%%%%%%%%%%%%%%%%%%%%%%%%%%%%%%%%
\subsection{Experiment workflow}
\label{subsec:experiment_workflow}
\subsubsection{Experiments using different connectivity maps}

Enter the folder named test in the nassc\_code folder and you can see there are total five folders in the folder test. Folder test\_CouplingMap\_FullyConnected, folder test\_CouplingMap\_linear, folder test\_CouplingMap\_grid and folder test\_CouplingMap\_montreal contains differents script to run benchmarks with four different coupling maps. And folder yaml\_file contains YAML files which are used to set different configurations when running the benchmark. For example, if you want to run benchmark grover\_n4 using the coupling map from device \texttt{ibmq\_montreal}, you can enter the folder test\_CouplingMap\_montreal and run the benchmark grover\_n4 by the following commands:

\$ python ./run\_benchmark.py ../yaml\_file/grover\_n4.yaml

If you want to collect all the result from all the benchmarks, go back to the directory /nassc\_code/test and use the script generate\_raw\_data.sh:

\$ ./generate\_raw\_data.sh

Running the above script generate\_raw\_data.sh will take about 10 hours to finish. After the script generate\_raw\_data.sh finishes running and getting all the result, run the following python script to generate the cnot\_table\_using\_montreal\_map.csv which corresponding to the results in Table~\ref{table:result_montreal}:

\$ python generate\_cnot\_table\_using\_montreal\_map.py

Run the following python script to generate depth\_table\_using\_montreal\_map.csv which corresponding to the results in Table~\ref{table:depth_montreal_result}:

\$ python generate\_depth\_table\_using\_montreal\_map.py

Run the following python script to generate cnot\_table\_using\_linear\_map.csv which corresponding to the results in Table~\ref{table:lnn_results}:

\$ python generate\_cnot\_table\_using\_linear\_map.py

Run the following python script to generate cnot\_table\_using\_grid\_map.csv which corresponding to the results in Table~\ref{table:2d_results}:

\$ python generate\_cnot\_table\_using\_grid\_map.py \\

\subsubsection{Experiments using noise model from real device ibmq\_montreal }
\label{subsubsec:experiment_real_montreal}

If you do not have access to the real IBM quantum device \texttt{ibmq\_montreal}, skip this subsection's experiments and go to the next subsection~\ref{subsubsec:experiment_fake_montreal}.

Before running the experiments, available IBMQ providers need to be set in the file of /nassc\_code/hamap/hardware/IBMQHardwareArchitecture.py by modifying the line 122 to use IBMQ providers that are available to you.

All the YAML files in the directory /nassc\_code/test\_HardwareAware/yaml\_file specify the configuration of running benchmarks, and all these YAML files need to be modified in order to use the IBMQ providers that are available to you. You can specify the IBMQ providers by modifying line 2 of all the YAML files. 

After finishing the configuration of IBMQ provider, go back to the folder test\_HardwareAware:

run the script generate\_raw\_data.sh to generate the benchmark results:

\$ ./generate\_raw\_data.sh

After the above command finishes, run the following python scripts to generate the cnot\_compare.pdf and SuccessRate\_compare.pdf which corresponding to the Figure~\ref{fig:noise-comparison} in our paper:

\$ python generate\_cnot\_table.py

\$ python generate\_SuccessRate\_table.py \\

\subsubsection{Experiments using noise model from fake ibmq\_montreal}
\label{subsubsec:experiment_fake_montreal}

The experiments in this subsection is provided as substitutions for the experiments in the above subsection~\ref{subsubsec:experiment_real_montreal} if you do not have access to a real IBM quantum device \texttt{ibmq\_montreal}.

Go to the directory /nassc\_code/test\_HardwareAware\_
strategy\_using\_backed\_up\_data, run the following command to get the benchmark results:

\$ ./generate\_raw\_data.sh

After the above command finishes, run the following python scripts to generate the cnot\_compare.pdf and SuccessRate\_compare.pdf which corresponding to the Figure~\ref{fig:noise-comparison} in our paper:

\$ python generate\_cnot\_table.py

\$ python generate\_SuccessRate\_table.py

However, the SuccessRate\_compare.pdf generated will be slightly different from the Figure~\ref{fig:successrate-comparison} in our paper, because the above result are generated using the noise model from FakeMontreal while Figure~\ref{fig:successrate-comparison} is generated using the noise model from real device \texttt{ibmq\_montreal}. 
%%%%%%%%%%%%%%%%%%%%%%%%%%%%%%%%%%%%%%%%%%%%%%%%%%%%%%%%%%%%%%%%%%%%%
\subsection{Evaluation and expected results}

After running the scripts mentioned in the experiments workflow, csv files and pdf files are generated, which corresponding to the result of Section~\ref{sec:evaluation}(Table~\ref{table:result_montreal}, Table~\ref{table:depth_montreal_result}, Table~\ref{table:lnn_results}, Table~\ref{table:2d_results}, Figure~\ref{fig:noise-comparison}).  

%%%%%%%%%%%%%%%%%%%%%%%%%%%%%%%%%%%%%%%%%%%%%%%%%%%%%%%%%%%%%%%%%%%%%
\subsection{Experiment customization}

The configuration in the YAML files can be modified to run different benchmarks with various configuration.

%%%%%%%%%%%%%%%%%%%%%%%%%%%%%%%%%%%%%%%%%%%%%%%%%%%%%%%%%%%%%%%%%%%%%
\subsection{Notes}
In the above Section~\ref{subsec:experiment_workflow}, experiments in subsection~\ref{subsubsec:experiment_fake_montreal} are provided as substitutions for experiments in subsection~\ref{subsubsec:experiment_real_montreal}. If you do not have access to the IBMQ quantum device \texttt{ibmq\_montreal}, you can skip the experiments in subsection~\ref{subsubsec:experiment_real_montreal} and perform the experiments in subsection~\ref{subsubsec:experiment_fake_montreal}.
%%%%%%%%%%%%%%%%%%%%%%%%%%%%%%%%%%%%%%%%%%%%%%%%%%%%%%%%%%%%%%%%%%%%%
\subsection{Methodology}

Submission, reviewing and badging methodology:

\begin{itemize}
  \item \url{https://www.acm.org/publications/policies/artifact-review-badging}
  \item \url{http://cTuning.org/ae/submission-20201122.html}
  \item \url{http://cTuning.org/ae/reviewing-20201122.html}
\end{itemize}

\end{document}